\documentclass[letter,11pt]{article}

\usepackage{fullpage}

\usepackage{amssymb}
\usepackage{amsmath}

\usepackage{wrapfig}
\usepackage{subcaption}
\usepackage{framed}
\usepackage{footmisc}
\usepackage[round,authoryear]{natbib}
\usepackage{authblk}
\usepackage{colortbl}
\definecolor{webgreen}{rgb}{0,0.4,0}
\definecolor{webbrown}{rgb}{0.6,0,0}
\definecolor{purple}{rgb}{0.5,0,0.25}
\definecolor{darkblue}{rgb}{0,0,0.7}
\definecolor{darkred}{rgb}{0.7,0,0}
\usepackage{hyperref}
\hypersetup{colorlinks,citecolor=darkblue,filecolor=black,linkcolor=darkred,urlcolor=webgreen,pdfpagemode=None,
pdfstartview=FitH}

\newcommand{\ignore}[1]{}

\newcommand{\given}{\ | \ }

\newtheorem{lemma}{{\sc Lemma}}

\newtheorem{theorem}{{\sc Theorem}}
\newtheorem{definition}{{\sc Definition}}

\newtheorem{claim}{{\sc Claim}}

\newtheorem{assumption}{{\sc Assumption}}

\usepackage{cleveref}
\crefname{claim}{claim}{claims}
\crefname{fact}{fact}{facts}
\crefname{algorithm}{algorithm}{algorithms}
\crefname{observation}{observation}{observations}
\crefname{equation}{equation}{equations}
\crefname{assumption}{assumption}{assumptions}

\newenvironment{proof}{\noindent {\em Proof\/}:\enspace}
{\hfill $\blacksquare{}$ \medskip \\}

\usepackage{titlesec}

\titlespacing*{\section}{0pt}{6pt plus 2pt minus 2pt}{6pt plus 2pt minus 2pt}
\titlespacing*{\subsection}{0pt}{6pt plus 2pt minus 2pt}{6pt plus 2pt minus 2pt}
\titlespacing*{\subsubsection}{0pt}{6pt plus 2pt minus 2pt}{6pt plus 2pt minus 2pt}

\DeclareMathOperator*{\argmin}{\arg\!\min}
\DeclareMathOperator*{\argmax}{\arg\!\max}

\newcommand{\E}{\mathbb{E}}

\usepackage{pgf}
\usepackage{verbatim}
\usepackage{enumerate}
\usepackage{amssymb}
\usepackage{mathrsfs}
\usepackage{algorithm}
\usepackage[noend]{algorithmic}

\usepackage{multirow,bigdelim}
\usepackage{cancel}
\usepackage{resizegather}

\usepackage{url}

\usepackage{xspace}

\usepackage{nicefrac}

\usepackage{enumitem}

\allowdisplaybreaks
\usepackage[margin=1in]{geometry}

\newcommand{\mechabbrv}{\text{\tt PEQA}}
\newcommand{\mech}{{\bf \texttt{P}}eer {\bf \texttt{E}}valuation with {\bf \texttt{Q}}uality {\bf \texttt{A}}ssurance}

\usepackage{cleveref}
\crefname{claim}{claim}{claims}
\crefname{fact}{fact}{facts}
\crefname{algorithm}{algorithm}{algorithms}
\crefname{observation}{observation}{observations}
\crefname{equation}{equation}{equations}
\crefname{assumption}{assumption}{assumptions}
\crefname{hypothesis}{hypothesis}{hypotheses}
\usepackage{enumerate}
\usepackage{subcaption}
\usepackage[noend]{algorithmic}
\usepackage{resizegather}
\usepackage{enumitem}
\usepackage{multirow,bigdelim}
\usepackage{cancel}
\usepackage{framed}
\usepackage{wrapfig}
\usepackage{tfrupee}

\newcommand{\independent}{\perp \!\!\! \perp}

\newtheorem{hypothesis}{Hypothesis}

\title{\bf Removing Bias and Incentivizing Precision in Peer-grading}

\author[1]{Anujit Chakraborty}
\author[2]{Jatin Jindal}
\author[3]{Swaprava Nath}

\affil[1]{\small University of California Davis, \texttt{chakraborty@ucdavis.edu}}
\affil[2]{\small Zamstars, \texttt{jatinjindal369@gmail.com}}
\affil[3]{\small Indian Institute of Technology Kanpur, \texttt{swaprava@iitk.ac.in}}

\date{}

\begin{document}
\maketitle

\begin{abstract}
\noindent
We study peer-grading with competitive graders who enjoy a higher utility when their peers get lower scores. 
We propose a new mechanism, \mechabbrv, that incentivizes such graders through a {\em score-assignment rule} which aggregates the final score from multiple peer-evaluations, and a {\em grading performance score} which rewards performance in the peer-grading exercise. 
\mechabbrv\ makes grader-bias irrelevant. Additionally, under \mechabbrv, a peer-grader's utility increases monotonically with the reliability of her grading, irrespective of her competitiveness and how her co-graders act. In a reasonably general class of score assignment rules, \mechabbrv\ uniquely satisfies this utility-reliability monotonicity. When grading is costly and costs are private information, a modified version of \mechabbrv\ implements the socially optimal effort-choices in an equilibrium of the peer-evaluation game. Data from our classroom experiments confirm our theoretical assumptions and show that \mechabbrv\ outperforms the popular {\em median} mechanism.

\end{abstract}
 

\section{Introduction}
\label{sec:intro}

A peer-evaluation process aggregates assessments from peers to judge the quality of submitted work. Scientific communities use  peer-evaluation for reviewing the quality of articles and grant proposals~\citep{campanario1998peer}. Coursera and EdX that offer Massive Open Online Courses (MOOCs) to 94 million learners\footnote{Numbers from \href{https://about.coursera.org/press/wp-content/uploads/2020/09/Coursera-Impact-Report-2020.pdf}{Coursera}'s and \href{https://www.edx.org/assets/2020-impact-report-en.pdf}{EdX}'s 2020 impact reports.}, use peer-grading to evaluate submitted assignments. Many in-person classes are also adopting it and its growing popularity can be explained by the following three reasons. First, it simplifies and accelerates the evaluation and grading process. Second, it improves learning outcomes of the participating students \citep{sadler2006impact}. Third, it easily scales to large classes.


When students are evaluated on a curve, students naturally care about their relative performance vis-à-vis peers. Even when evaluated on an absolute grading scale, students care about their relative performance due to the role it plays in admission into jobs or higher studies. This creates perverse incentives for peer-graders. \cite{strong2004self} find that peer-graders give consistently biased grades in peer-grading schemes. In an anonymous survey that we ran on the students of a reputed technical institute in India, 49\% of the 549 respondents expected that their fellow students would grade aggressively to reduce the scores of others, and thereby try to improve their relative class-ranking. 

We study the problem of incentivizing competitive and strategic peer-graders. In our model, students write an exam and then peer-grade each others' exams. Thus, every student has dual roles: (i) the student role, where she writes an exam that gets evaluated, and (ii) the grader role, where she evaluates others. Their total course-score is the sum of their own exam score (aggregated from peer-reports) and a score based on their peer-grading performance. To model competitive students, we assume that their utility is linearly increasing in their total course-score and linearly decreasing in their peers' total course-scores. 

To model strategic grading, we adapt the $\mathbf{PG}_1$ statistical peer-grading model of \citet{piech2013tuned} to a strategic environment.\footnote{This is a widely used statistical model for peer evaluation in MOOCs.}
Our model assumes that each paper being peer-graded has a true score. Peer-graders choose the reliability (inverse of variance) of the independent, noisy signals that they observe about the true score. Choosing higher reliability results in observing a more accurate signal. 
Graders can then decide to add a bias to their observed signal while reporting their assessment. Graders who care about their relative success within peers might purposefully bias their evaluations. They may also choose to receive
less reliable signals.

What is the set of desiderata one could ask for a mechanism in this setup? At a minimum, the mechanism should be able to overcome the perverse competitive incentives of biased or unreliable grading. To simplify, we initially assume that more reliable grading (lower variance) does not come at an extra cost to the peer-grader. 

We propose a new mechanism, \mech\ (\mechabbrv), that ensures that (\Cref{thm:bi-rm}):
\begin{itemize}[noitemsep,leftmargin=*,topsep=0pt,parsep=0pt,partopsep=0pt]
    \item Assigned scores and grader's utility are {\em bias-insensitive} (defined in \Cref{def:bi}).
    \item Higher reliability ensures monotonically higher utility to the grader, despite her competitiveness and actions of her co-graders. ({\em reliability monotonicity}, \Cref{def:rm}). 
    \item \mechabbrv\ uniquely satisfies the monotonic reliability-utility relation within a moderately general class of mechanisms (\Cref{thm:grade-unique}).
\end{itemize}   

In \Cref{sec:costlyeffort}, we address if \mechabbrv\ satisfies the more ambitious desiderata of implementing a ``preferred level" of grading among competitive graders, while accounting for the cost of grading-effort. We assume that students face an additional disutility (cost) from grading that increases with their reliability. How much effort should one ask students to exert? Reliability is desirable, but it might be prohibitively costly for students to spend all their time on grading! 
We define the {\em net social utility} (\Cref{eq:social-utility}) from the game as the difference between the social benefit of high reliability and the aggregate cost of effort. 
%
Under this setup, we show that:
\begin{itemize}[noitemsep,leftmargin=*,topsep=0pt,parsep=0pt,partopsep=0pt]
    \item A modified version of \mechabbrv\ implements Nash equilibria of the peer-grading game (with private costs) in which individuals spend the socially-optimal level of effort (\Cref{thm:costly}).
    \item The modified \mechabbrv\ maintains the same ranking among the students as the original \mechabbrv\ (\Cref{lemma:order-inv}).
\end{itemize}

How does the mechanism \mechabbrv\ work? A small subset of the total number of papers (called \emph{probes}) is evaluated by the teaching staff. Each grader is assigned $K>2$ papers (with $K/2$ probes) and they never grade their own paper. The peer-graders cannot tell apart the probes from the non-probes. \mechabbrv\ compares the grader's and the teaching staff's evaluations of the probes to estimate each grader's bias and reliability.\footnote{In the peer-review process, associate editors and editors can independently evaluate a subset of refereed manuscripts to learn if the assigned referees made a fair assessment on the papers.} 
This requires two identifying assumptions: that the teaching staff can observe the true scores on the probe papers, and, that the graders grade identically on probes and non-probes.
The estimated grader-bias is subtracted from the peer-reports to de-bias the reports. \mechabbrv's score-assignment function assigns a weighted average of the de-biased grader-reports, with the weights being the inverse square-root of the estimated grader-variance. Thus, reports from high variance graders play a smaller role in the finally assigned score. 

\mechabbrv\ derives grading performance scores from its score-assignment function. The performance scores resemble the Vickrey-Clarke-Groves (VCG) transfer~\citep{Vick61,Clar71,Grov73}.

\begin{figure}[h!]
 \centering
 \includegraphics[width=0.9\linewidth]{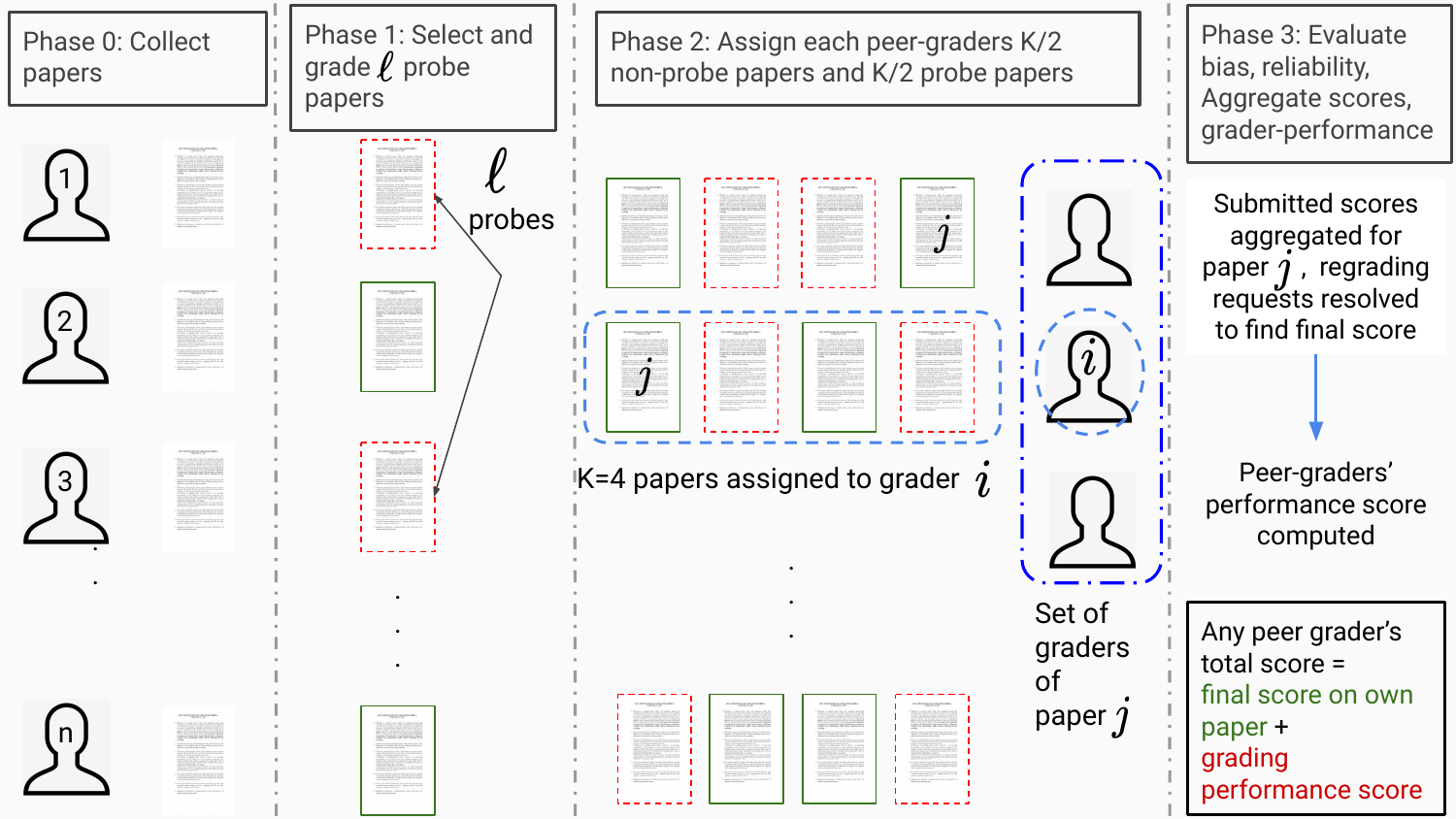}
 \caption{Schematic diagram of the \mechabbrv\ mechanism decomposed into {\em four} phases. A typical non-probe paper is denoted by $j$ here.}
 \label{fig:pg}
\end{figure}

We allow students to raise regrading requests after seeing their score. The teaching staff regrade such papers and assign them the true score. We assume that such requests are raised only when the student knows that her initially assigned score was lower than the true score.\footnote{When a manuscript is rejected based on an objectively incorrect referee-report, authors can similarly appeal to the editor. But, this is far less common.} The schematic diagram of the stages of \mechabbrv\ is shown in \Cref{fig:pg}.

To test some of our baseline assumptions and to see how easily our mechanism could be implemented in practice, we ran classroom experiments (\Cref{sec:human}). Students enrolled in a computing course were asked to peer-grade a weekly class-quiz.
The scores assigned under \mechabbrv\ were remarkably accurate, and only $1$ out of $41$ had a wrong score. 

Results from our \mechabbrv\ sessions (\Cref{tab:Results-from-Median,tab:Results-from-TRUEQA,tab:Comparison-between-Median-TRUPEQA}) confirm two of our assumptions. 
\begin{enumerate}[noitemsep,topsep=0pt,parsep=0pt,partopsep=0pt]
 \item The bias and variance were indeed identical across probes and non-probes: subjects were not able to discern one from the other (\Cref{hyp:bias-equal}).
 \item Grade-manipulations, whenever present, reduced scores instead of inflating scores. This rejects the existence of collusive (i.e., the opposite of competitive) graders (\Cref{hyp:bias-sign}).
\end{enumerate}
We ran a second competitive session under a Median mechanism, which is currently the most popular mechanism used in MOOCs.\footnote{Peer reports are aggregated through the median score, as reported on \href{https://learner.coursera.help/hc/en-us/articles/208279946-Getting-and-viewing-grades-for-peer-reviewed-assignments}{Coursera} and \href{https://support.edx.org/hc/en-us/articles/360000192027-How-are-peer-assessment-scores-calculated-rules}{EdX} websites.}
In our experiments, \mechabbrv\ mechanism outperformed Median mechanism in terms of allocating accurate final scores (\Cref{hyp:better-final-score}). These differences were statistically significant. We have also developed a peer-grading platform {\tt SwaGrader} (\url{swagrader.cse.iitk.ac.in}) which uses \mechabbrv\ as the main peer-grading algorithm and is being tested by instructors and students within the Indian Institute of Technology Kanpur.

\subsection{Related Work}
\label{sec:literature}

The existing research on peer-evaluation mechanisms can be broadly divided into three strands. The first strand of literature abstracts away from any strategic motives of the peer-evaluators. Instead of providing a mechanism to incentivize strategic evaluators, they propose how the grader reports could be aggregated efficiently
\citep{hamer2005method,cho2007scaffolded,piech2013tuned,shah2013case,pare2008peering,kulkarni2014scaling,de2014crowdgrader,raman2014methods,caragiannis2015aggregating,wright2015mechanical}. 

The second strand of literature is based on {\em peer-prediction} approaches. These mechanisms incentivize {\em coordination} on similar evaluation reports by punishing evaluations that don't match each other. Thus, they do not necessarily incentivize {\em accuracy}~\citep{prelec2004bayesian,miller2005eliciting,jurca2009mechanisms,faltings2012eliciting,witkowski2013dwelling,dasgupta2013crowdsourced,witkowski2013learning,waggoner2014output,shnayder2016informed}. Any such mechanism introduces uninformative equilibria alongside the truth-telling one \citep{jurca2009mechanisms,waggoner2014output}.\footnote{In particular, when the information is costly to obtain, it is generally easier for the agents to resort to coordinating on an uninformative low-effort equilibrium.} 
More recent developments make the truthful equilibrium Pareto dominant,
 i.e., the truthful equilibrium is (weakly) more rewarding to every agent than any other equilibrium \citep{dasgupta2013crowdsourced,witkowski2013learning,kamble2015truth,radanovic2015incentives,shnayder2016informed}. 

The final strand consists of {\em hybrid} approaches where the true quality of some of the peer-assessed material can be found, for e.g, via evaluating a part of the materials by the mechanism designer (teaching staff in case of MOOCs) herself. Graders are then rewarded for agreement with the designer-agreed report \citep{jurca2005enforcing,dasgupta2013crowdsourced,gao2016incentivizing}. 
Our mechanism also utilizes the feature that the true scores on a small subset of assignments can be revealed at a small cost. However, additionally, we address new and practical features of the peer-grading probem: we allow for competitive graders, we solve the efficient allocation problem under costly grading, and we allow regrading requests.

\cite{alon2011sum} and \cite{holzman2013impartial} study situations where peers have to choose a subset amongst themselves for a reward. The challenge here is to incentivize the peers to reveal their private information unselfishly. In particular, the goal is to guarantee that what peers report does not affect their chances of winning or getting selected. In these settings, there is no need to incentivize peers to gather information that is `objective' (e.g., true score on an exam) and verifiable at a cost. There is also no need to ensure that peers enjoy higher utility when their gathered information is more precise. Finally, peers are purely selfish: they do not care about who wins in case they do not win themselves. Thus, by debriding the reports from personal winning chances, the mechanism makes the peers indifferent between all reports.

\cite{cai2015optimum} consider a setting where data-sources (e.g., human labelers) can be paid monetarily to get their estimation of $f(x_i)$ at points $x_i$ allocated to them. The end goal is to estimate an exogenously provided $f$ using a given estimator $\hat{f}$. 
Data-sources can observe a noisy version of $f(x_i)$ with the noise decreasing in their effort, and they maximize the difference between the payment and the cost of the effort. They show that under their VCG-like payment mechanism and the assumption of a ``well-behaved" $\hat{f}$, the dominant strategy for a data-source is to reveal its observation correctly and always participate in the data-providing exercise. \cite{cai2015optimum}'s data-sources naturally have no competitive preferences, like our graders do. We also propose the optimal estimator $\hat{f}$ to use on the observed data, which they do not.

\section{Peer-grading Mechanism}
\label{sec:mechanisms}

\subsection{Definition} 
\label{sec:pg-defn}

Each subject $i \in  N=\{1,\ldots,n\} $ has written an exam, and is also a participant in the peer-grading process. Thus $N=\{1,\ldots,n\}$ represents both the set of papers to be graded and  the set of graders. We use $i$ as the index for a grader and $j$ as the index for a paper. For simplicity of exposition, we assume that each paper has only {\em one} question for evaluation. 

Our mechanism would instruct the teaching staff to evaluate a fixed number $\ell (<< n)$ of these papers so that their true grades are known. These papers are called the {\em probe} papers. Let $G(j)$ denote the set of peer-graders of paper $j$ and $G^{-1}(i) := \{k \in N: i \in G(k)\}$ denote the set of papers assigned to evaluator $i$. The set $P_i \subset G^{-1}(i) $ and $NP_i = G^{-1}(i) \setminus P_i$ denotes respectively the probe and non-probe papers assigned to $i$.  Both true and reported scores belong to 
$\mathbb{R}$. 
The co-graders of individual $i$ are $CG_i = \cup_{j \in NP_i} G(j) \setminus \{i\}$. We assume that the co-graders of $i$ grade at least one common non-probe paper with $i$.

\smallskip
\noindent
Assuming that peer-reported scores are real numbers, a {\em peer-grading mechanism} $M$ is the tuple $\langle G, \mathbf{r}, \mathbf{t} \rangle$, where 
\begin{itemize}[noitemsep,leftmargin=*,topsep=0pt,parsep=0pt,partopsep=0pt]
 \item $G$ is the {\em assignment} function $G: N \to 2^N$ that maps papers to graders. 
 \item $\mathbf{r} : \times_{j\in N} \mathbb{R}^{G(j)} \to \mathbb{R}^n$ is 
the {\em score-assignment} function, where the $j$th 
component $r_j (\cdot)$ is the function assigning the final 
score of paper $j$ based on the scores reported by $G(j)$.
 \item $\mathbf{t} : \times_{i\in N} \mathbb{R}^{G^{-1}(i)} \to \mathbb{R}^n$ is 
the {\em peer-grading performance score} function, where the $i$th 
component $t_i (\cdot)$ is the function that yields the  
peer-grading performance score to grader $i$.
\end{itemize}
Since every student $i$ has dual roles in peer-grading as explained in \Cref{sec:intro}, $r_i$ and $t_i$ are the mechanism-assigned scores corresponding to her student and grader roles. For example, in a course that has 80 points on the exam and 20 points on peer grading performance, a student might score $r_i=60$ and $t_i=15$ on those two respectively. Her total course-score would be $75$ out of $100$.

\subsection{Model of the True and Reported Scores} 
\label{sec:model}

We generalize the $\mathbf{PG}_1$ model of true score, bias, and, reliability \citep{piech2013tuned} to a strategic environment. We make two major changes. First, we replace their assumptions of normality with a general distribution $\mathcal{F}(\cdot)$ with a support of $(-\infty,\infty)$ and a {\em differentiable} density function $f(\cdot)$.
We use $\mathcal{F}(\mu,1/\gamma)$ for such a distribution with mean $\mu$ and variance $1/\gamma$. Second, instead of assuming that bias and reliability are drawn randomly and independently from Normal and Gamma distributions respectively, we make each a strategic choice by the peer-graders. Subject to these changes, the following features in our model resemble the $\mathbf{PG}_1$ model.
 
\begin{figure}[h!]
 \centering
 \includegraphics[width=0.7\linewidth]{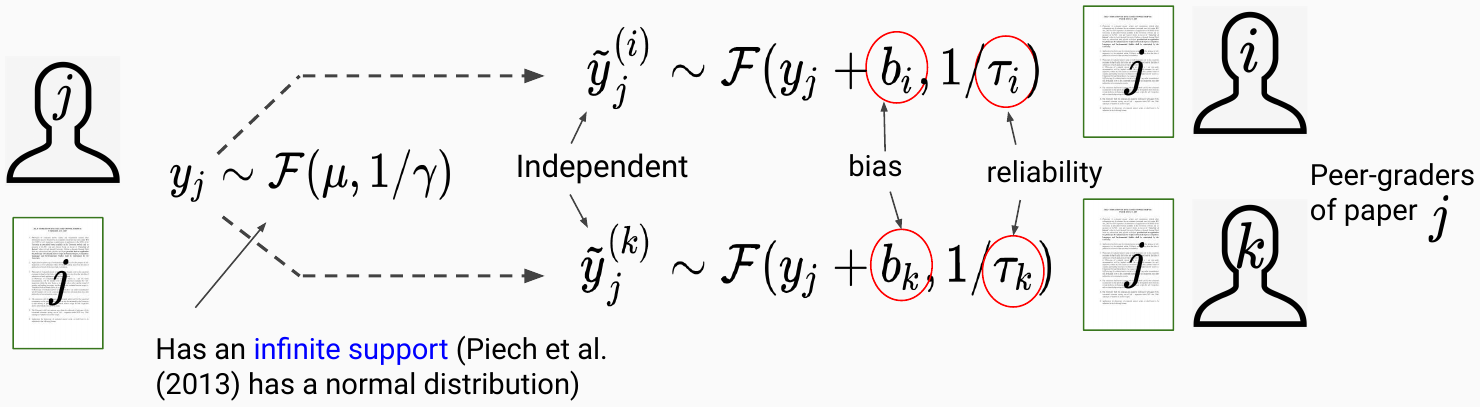}
 \caption{Peer-reports' generation process.}
 \label{fig:pg_model}
\end{figure}

\begin{itemize}[noitemsep,leftmargin=*,topsep=0pt,parsep=0pt,partopsep=0pt]
  \item The true score $y_j$ for paper $j$ is distributed as $\mathcal{F}(\mu,1/\gamma)$, for all $j \in N$. This distribution is known from historical data of past examinations. 
 \item Peer-graders do not see $y_j$ but after they choose their reliability $\tau_i$, they observe an independent draw from $\mathcal{F}(y_{j},1/\tau_{i})$. Higher is $1/\tau_{i}$, noisier is the draw.
 \item Graders then add a bias $b_i$ to the signal before reporting. $\tilde{y}_{j}^{(i)}$ is the reported score of paper $j$ by grader $i$. Conditional on the true score $y_j$, it is distributed as $f(\tilde{y}_{j}^{(i)}|y_j) \sim \mathcal{F}(y_{j}+b_{i},1/\tau_{i})$, where $b_i$ and $\tau_i$ are called the {\em bias} and {\em reliability} of $i$ respectively. 
 \item Independence across exams and graders: The conditional distributions of $i$ and $k$'s reported scores on exams ${j_1}$ and ${j_2}$ are independent. Thus, $f(\tilde{y}_{j_1}^{(i)}|{y}_{j_1}) \independent f(\tilde{y}_{j_2}^{(k)}|{y}_{j_2})$, for all $i,k, j_1,j_2,{y}_{j_1},{y}_{j_2}$ such that $i=k$ and $j_1=j_2$ aren't true simultaneously. Thus, even for the same grader, the signals from different exams are conditionally independent. And, even for the same exam, the signals received by different graders are  conditionally independent.
 \item We have used the same distribution $\mathcal{F}$ for both the true scores $y_j$ and the score observed by the grader $i$, i.e., $\tilde{y}_{j}^{(i)}$, to keep the model similar to $\mathbf{PG}_1$. However, this is not critical to our results. In particular, (a)~we can have two different distributions for these two sets of random variables, and (b)~the distribution of the observed score $\tilde{y}_{j}^{(i)}$ can vary with $i$. None of these will affect the main conclusions of this paper. 
 \end{itemize}
The dynamics of the grading process is shown in \Cref{fig:pg_model}. 
Reliability is defined as the inverse of noise variance. Bias originates from a strategic manipulation or from non-strategic (generous or strict) grading-habits. 
In this paper, we would assume that the grader chooses her bias and reliability. 

We assume that a grader grades all papers (probes and non-probes) with the same bias and reliability. This assumption is natural if the graders cannot tell the probes from the non-probes. We find support for this assumption in our experimental sessions. Bias and reliability are indeed identical across probes and non-probes.
We use the shorthand $\theta_i = (b_i, \tau_i) \in \mathbb{R} \times \mathbb{R}_{\geqslant 0}$ to denote grader $i$'s strategic choices.

\subsection{Other primitives of our mechanism}
\label{sec:primitives}
We have already defined a general peer-grading mechanism in \Cref{sec:pg-defn}. In this section, we fine-tune the $\langle G, \mathbf{r}, \mathbf{t} \rangle$ functions for our proposed mechanism. 
\subsubsection*{Paper assignment rule $G^*(\cdot)$} 
Every paper is graded by at least one grader, and every grader grades at least two probe and one non-probe papers. Thus, (a)~$G^*(j) \neq \emptyset$ and $j \notin G^*(j), \ \forall j \in N$, (b)~$|P_i| \geqslant 2, \ \forall i \in N$, and (c)~$NP_i \neq \emptyset, \ \forall i \in N$. The graders know the proportion of probe and non-probe papers assigned to them, but cannot tell them apart.  

\subsubsection*{Grade assignment and performance scores} The mechanism compares the peer-graded scores ($ \tilde{y}_{j}^{(i)}$)  with  true scores ($y_j$) on the probe papers $P_i$, to statistically estimate the error parameters $\hat{\theta}_i = (\hat{b}_i, \hat{\tau}_i) \in \mathbb{R} \times \mathbb{R}_{\geqslant 0}$ of each grader $i$. We have $\hat{b}_i = \frac {\sum_{j \in P_i} ( \tilde{y}_{j}^{(i)} - {y}_{j} ) }{|P_i|}$ and $\hat{\tau}_i = \frac{|P_i|-1}{\sum_{j \in P_i} (\tilde{y}_{j}^{(i)}-(y_{j}+\hat b_{i}))^2}$. 
The estimated parameters are used in assigning performance-scores to papers and performance scores to peer-graders.

\begin{definition}[Score and Reward]
 \label{def:score-welfare}
 We define the score-assignment rule and the social reward as follows.
 
\begin{itemize}[noitemsep,leftmargin=*,topsep=0pt,parsep=0pt,partopsep=0pt]
  \item The score-assignment function $\mathbf{r} = (r_j : j \in N)$ is {\em inverse standard-deviation weighted de-biased mean (ISWDM)} if for every non-probe paper $j$, it assigns
  \begin{equation}
   \label{eq:r}
   r_{j}^{*}(\mathbf{\tilde{y}}^{G(j)}_{j}, \hat{\boldsymbol{\theta}}_{G(j)}) = \frac{\sqrt{\gamma}\mu+\sum_{i \in G(j)}\sqrt{\hat{\tau}_{i}}(\tilde{y}_{j}^{(i)}-\hat{b}_{i})}{\sqrt{\gamma}+\sum_{i\in G(j)}\sqrt{\hat{\tau}_{i}}},
  \end{equation}
  where $\tilde{y}_{j}^{(i)}$ is the evaluation by the $i$th peer-grader and $(\hat{b}_{i},\hat{\tau}_{i})$ are her estimated parameters. Score $\mathbf{r}^*$ assigns the instructor-verified grade on every probe paper. 
  \item The {\em social reward} for paper $j$, at a score $r_{j}^{*}$  and true score $y_j$, is 
  \begin{equation}
   \label{eq:W}
   W_j^{*}(\mathbf{\tilde{y}}^{G(j)}_{j}, \hat{\boldsymbol{\theta}}_{G(j)}, y_j) = R(r_j^*(\mathbf{\tilde{y}}^{G(j)}_{j}, \hat{\boldsymbol{\theta}}_{G(j)}), y_j),
  \end{equation}
where $\mathbf{\tilde{y}}^{G(j)}_{j}$ is the vector of 
peer-evaluated scores reported on paper 
$j$, $\hat{\boldsymbol{\theta}}_{G(j)}$ is the vector of 
evaluated error-parameters for the relevant graders $G(j)$, 
and $R:\mathbb{R}^2 \to \mathbb{R}$ is a reward function that measures the closeness of the true score $y_j$ and the given score $r_j^*$.

Formally, $R(x_1,y_1) < R(x_2,y_2)$ if $|x_1-y_1| > |x_2-y_2|$ for all $x_1,x_2,y_1,y_2 \in \mathbb{R}$. We assume that $R(x,x)=0\geqslant R(x,y)=R(y,x)$ for all $x,y \in \mathbb{R}$.
One example of such a function would be $R(x,y)=-(x-y)^2$, 
which calculates the squared error in assigned scores. 

  \item The {\em social reward} at a score $r_{j}^{*}$ for paper $j$ {\em without grader $i$} when the true score is $y_j$ is denoted by $W^{(-i)*}_j=W_j^*(\mathbf{\tilde{y}}^{G(j) \setminus \{i\}}_{j}, \hat{\boldsymbol{\theta}}_{G(j) \setminus \{i\}}, y_j)$
  where $W^*_j(\cdot)$ is defined as above.
 \end{itemize}
 The parameters $\gamma, \mu, b_i$, and $\tau_i$ are as given by the $\mathbf{PG}_1$ model of \citet{piech2013tuned} (see \Cref{sec:model}).
\end{definition}
We will use the shorthands $W_j^{*}$ and $W_j^{(-i)*}$ for the social rewards with and without agent $i$ respectively when the arguments of such functions are clear from the context.

The ISWDM score-assignment function takes a {\em weighted average} of the prior mean $\mu$ and the de-biased (subtracting the estimated bias from the reported scores) reported scores. De-biasing ensures that the biases of the graders do not affect the finally assigned grade. The weight is chosen to be the square-root of reliability, which is the inverse of the variance for that grader. Higher the estimated reliability, higher is the weight on a grader.

Without incentive concerns, a statistician would have suggested a score-assignment function that would minimize the expected squared distance between the assigned score and true score on exam $j$, conditional on the true bias and variance parameters. Then, those true parameters could be approximated by the estimated bias and variance. In \Cref{eqn:erm_est} of \Cref{sec:erm-calculation}, we show that under the strong distribution-assumptions of \citet{piech2013tuned}, such a score-assigment function on exam $j$ would come from the class of {\em weighted average (WA) score-assignment functions}:
\begin{equation}
 \label{eq:r-general-k}
 r_j^\text{WA}(\mathbf{\tilde{y}}^{G(j)}_{j}, \hat{\boldsymbol{\theta}}_{G(j)}) = \frac{\lambda_0 \mu+\sum_{i \in G(j)}\lambda_i(\tilde{y}_{j}^{(i)}-\hat{b}_{i})}{\lambda_0+\sum_{i\in G(j)}\lambda_i}, 
\end{equation}
where $\lambda_0, \lambda_i \geqslant 0, \forall i \in N$, not all zero. In particular, the parameters turn out to be $\lambda_0 = \gamma$ and $\lambda_i = \hat{\tau}_{i}, \forall i \in N$ (note the diffence with $\lambda_i = \sqrt{\hat{\tau}_{i}}$ in \Cref{eq:r}; see \Cref{sec:erm-calculation} for details). 
Here, $\mu$ is the prior mean of all papers, and the term $(\tilde{y}_{j}^{(i)}-\hat{b}_{i})$ is the de-biased score on paper $j$ from grader $i$. This is indeed the {\em expected (social) reward maximizer (ERM)}, with the reward function $R(\cdot)$ being a negative quadratic function:
\begin{equation}
\label{eq:erm-r*}
 r_j^\text{ERM}(\mathbf{\tilde{y}}^{G(j)}_{j}, \hat{\boldsymbol{\theta}}_{G(j)}) \in
   \argmax_{x_{j} \in \mathbb{R}} \E_{y_{j} \given \mathbf{\tilde{y}}^{G(j)}_{j}; \hat{\boldsymbol{\theta}}_{G(j)}} R(x_{j}, y_{j}).
\end{equation} 
However, in  \Cref{thm:grade-unique} we will show that in the class of WA score-assignment functions, ISWDM uniquely satisfies certain
desirable properties. 
Even though ISWDM is not exactly the ERM, it does not compromise the expected social reward ($W_j$) much (see \Cref{sec:comparison2}).

\subsubsection*{Regrading Requests}
We consider peer-grading mechanisms that allow regrading requests. We assume that when a regrading request is raised, the instructor regrades the paper herself and assigns the true score on the paper. We also assume that the students know the true scores on their own papers and only raise a regrading request when they expect it to raise their score further.
\begin{assumption}
\label{ass:regrading}
 Student $j$ knows $y_j$ and raises a regrading request only if $r_j^* < y_j$.
\end{assumption}

In the next section, we lay down the peer-graders' incentive structure and the desirable properties of a mechanism.

\section{Incentives and Design Desiderata}
\label{sec:desiderata}


\paragraph{Individual Preferences.}  
We assume that every individual $i$ cares about (i)~her total score (sum of her exam score $r_i$ and peer-grading performance score $t_i$),
and (ii)~the total scores of the other individuals. To model a competitive grader who cares about her relative performance in the class, we assume that her utility is increasing in (i), weakly decreasing in (ii). This assumption is consistent with the \cite{strong2004self} finding that peer-graders give biased grades in peer-grading schemes. 

For agent $i$ in mechanism $M = \langle G, \mathbf{r}, \mathbf{t} \rangle$, the utility is given by
\begin{equation}
 \label{eq:utility}
 u_i^{M} = r_i + t_i -  \left( \sum_{j \in N \setminus \{i\}} w_{ij} \cdot (r_j +  t_j) \right),
\end{equation}
where $w_{ij} \geqslant 0$. 
%

In this section, we will assume that more reliable grading does not come at any extra cost for the peer-grader, and hence we exclude such a cost component from the utility expression. The objective here is to understand whether a peer-grading mechanism can reward more reliable grading monotonically, despite the presence of competitive preferences, and when increasing costs are not at play: we define the desirable properties accordingly.

One could have instead considered {\em costs of grading} to be increasing in reliability.  We do this in \Cref{sec:costlyeffort}, and the desiderata changes accordingly.


Note that a few uncertainties are resolved after grader $i$ chooses her decision variables $(b_i,\tau_i)$ and before $r^*$ and $t^*$ are computed by the mechanism: (i)~the scores are reported by grader $i$, $\tilde{y}_{j}^{(i)}$ for paper $j$, which is realized from $(\tilde{y}_{j}^{(i)}|y_j) \sim \mathcal{F}(y_{j}+b_{i},1/\tau_{i})$, (ii)~the decision variables $(b_k,\tau_k)$ are chosen by co-grader $k$ (i.e., the strategic uncertainty), (iii)~the true score $y_j$ on paper $j$ realizes,
and (iv)~the score on paper $j$ is reported by a co-grader $k$, which is realized from $(\tilde{y}_{j}^{(k)}|y_j) \sim \mathcal{F}(y_{j}+b_{k},1/\tau_{k})$. 
We define two desirable properties of peer-grading mechanisms. The properties consider the grader $i$'s expected utility from the choice of strategies she makes. 
All expectations are taken {\em only with respect to uncertainty (i)}, i.e., the distribution of $i$'s grade-evaluation process $(\tilde{y}_{j}^{(i)}|y_j) \sim \mathcal{F}(y_{j}+b_{i},1/\tau_{i})$. The properties hold for any ex-post realization of the other uncertainties (ii) to (iv), and there is no expectation taken on them. This is why both properties are defined as {\em ex-post}.

\begin{definition}[Ex-Post Bias Insensitivity (EPBI)]
 \label{def:bi}
A peer-grading mechanism $M = \langle G, \mathbf{r}, \mathbf{t} \rangle$ is {\em ex-post bias insensitive}, if 
the expected utility of participant $i$
is independent of the bias $b_i$. Bias independence holds irrespective of the biases and reliabilities of other graders $j\ne i$, the true score $y_j$, {\em and} reported scores of the other graders. Mathematically,
 \begin{align}
  \label{eq:dsbi}
 E_{ (\tilde{y}_{j}^{(i)}|y_j) \sim \mathcal{F}(y_{j}+b_{i},1/\tau_{i}) } u_i^{M} (\tilde{y}_j^{(i)}, \tilde{\mathbf{y}}_j^{(-i)}, y_j) 
 &=E_{(\tilde{y}_{j}^{(i)}|y_j) \sim \mathcal{F}(y_{j}+b'_{i},1/\tau_{i})} u_i^{M} (\tilde{y}_j^{(i)}, \tilde{\mathbf{y}}_j^{(-i)}, y_j), \nonumber \\
 &\qquad \forall \{\tilde{y}_j^{(k)}, b_k, \tau_k\}_{k \neq i}, \forall y_j, \tau_i, \forall j \in G^{-1}(i), \forall i \in N.
 \end{align}
\end{definition}

\begin{definition}[Ex-Post Reliability Monotonicity (EPRM)]
 \label{def:rm}
 A peer-grading mechanism $M =\langle G, \mathbf{r}, \mathbf{t} \rangle$ is {\em ex-post reliability monotone} if for every grader, the utility is {\em monotonically increasing} with her reliability, irrespective of the biases and reliabilities chosen by the other graders, the realizations of the true scores {\em and} the scores reported by the different graders. Mathematically,
 \begin{align}
  \label{eq:dsrm}
E_{(\tilde{y}_{j}^{(i)}|y_j) \sim \mathcal{F}(y_{j}+b_{i},1/\tau_{i})}  u_i^{M} (\tilde{y}_j^{(i)}, \tilde{\mathbf{y}}_j^{(-i)}, y_j)
  &> E_{(\tilde{y}_{j}^{(i)}|y_j) \sim \mathcal{F}(y_{j}+b_{i},1/\tau'_{i})}  u_i^{M} (\tilde{y}_j^{(i)}, \tilde{\mathbf{y}}_j^{(-i)}, y_j) , \nonumber \\
 &\forall \tau_i > \tau'_i, \forall \{\tilde{y}_j^{(k)}, b_k, \tau_k\}_{k \neq i}, \forall y_j, b_i, \forall j \in G^{-1}(i), \forall i \in N.
 \end{align}
\end{definition}
Both these properties are stronger than a {\em dominant strategy} version of the above definitions. The utility depends on the realizations of random variables like the true scores $\mathbf{y}$ and the reported scores $\mathbf{\tilde{y}}$. A dominant strategy definition will only require the (in)equalities to be satisfied after taking {\em interim} expectation over some relevant distribution over those variables. However, our {\em ex-post} properties require this to be satisfied for {\em every} realization of these random variables.

%
We are now in a position to present the central mechanism of this paper.

\section{The \mechabbrv\ mechanism}
\label{sec:mechanism}

\Cref{alg:trupeqa} shows the detailed steps of \mechabbrv.
%
\begin{algorithm}
\caption{\mechabbrv}
\label{alg:trupeqa}
 \begin{algorithmic}[1]
  \STATE Inputs: (1) the parameters $\mu$ and $\gamma$ of the priors on $y_j, \ \forall j \in N$, which is distributed as $\mathcal{F}(\mu,1/\gamma)$, (2) the reported scores $\mathbf{\tilde{y}}^N_{P}$ of the graders on the probe papers, and (3) reported scores $\mathbf{\tilde{y}}^N_{N \setminus P}$ on the non-probe papers.
  \STATE Set the probe set $P$ with $|P| = \ell$, a pre-determined constant $ \leqslant \frac{n}{\frac{K}{2}+1}$, where $K$ (even) is the number of papers assigned to each grader. 
  \STATE $G=G^*$: every grader $i \in N$ is assigned $\frac{K}{2}$ probe and $\frac{K}{2}$ non-probe papers, in such a way that every non-probe paper is assigned to at least $\frac{K}{2}$ and at most $\frac{K}{2}+1$ graders. This is always possible by assigning the $(n-\ell)$ non-probe papers to $(n-\ell)$ graders with each paper assigned to exactly $\frac{K}{2}$ graders. The rest $\ell$ graders can be assigned to the same $(n-\ell)$ papers arbitrarily such that these papers get at most one additional grader (since $\ell \frac{K}{2} \leqslant n-\ell$). Note that this is the reason $\ell$ cannot be larger than $n/(K/2 + 1)$. Ensure that a grader does not get her own paper for evaluation. \label{step:assignment}
  \STATE Estimate $\hat{b}_i, \hat{\tau}_i, \forall i \in N$ as given in \Cref{sec:primitives}.
  \STATE $\mathbf{r}$: the score of the paper $j$ is given by the ISWDM $\mathbf{r}^*$ (\Cref{eq:r}). \label{step:score-choice}
  \STATE At this stage, students may request for regrading. Instructor learns the correct grade $y_j$ for the papers which came for regrading. For other papers, $y_j = r_j^*$ is assumed. \label{step:regrading}
  \STATE $\mathbf{t}$: the performance score to grader $i$ for grading paper $j \in NP_i$ is given by $t_i^j = \alpha (W_j^{*} - W_j^{(-i)*})$, where $\alpha>0$ is a constant chosen at the designer's discretion. The total performance score to grader $i$ is therefore $t_i = \sum_{j \in NP_i} t_i^j$. \label{step:payment}
 \end{algorithmic}
\end{algorithm}
%
In short, the algorithm description specifies the three functions of a peer-grading mechanism $\langle G, \mathbf{r}, \mathbf{t} \rangle$ as defined in \Cref{sec:pg-defn}. 
The papers are assigned to the graders in a specific way. The assigned score on a paper is a weighted average (with appropriately chosen weights). Finally, the grading performance score is the {\em marginal contribution} of the grader towards the social reward. In the next section, we present our results on \mechabbrv.

\section{Properties of \mechabbrv}
\label{sec:properties}


Our first result shows that \mechabbrv\ satisfies both the properties mentioned in \Cref{sec:desiderata}, as long as the subjects care more about their own scores than others' scores. 

\begin{theorem}
 \label{thm:bi-rm}
 If $\sum_{k \in N\setminus\{i\}}w_{ik}\leqslant 1, \forall i\in N$, then \mechabbrv\ is EPBI and EPRM.
\end{theorem}
A direct consequence of this result is that a grader will have no incentive in putting a deliberate upward or downward bias in this competitive environment and also will find it in her interest to maximize her reliability. 
%

All the $r_i$ terms in the utility expression (\Cref{eq:utility}) would be replaced by $\max \{r_i,y_i\}$ due to \Cref{ass:regrading}, because the instructor is assumed to give the correct score $y_i$ when a regrading request is received. 

To make the proofs easily readable, we provide an intuition of the main ideas here. The complete details are available in \Cref{sec:proofs}.

The EPBI result is driven by how the score-assignment function de-biases the grades through the estimated grader bias. Though the bias estimates from probes are noisy, in expectation, they are correct and are identical across probes and non-probes. Thus grader $i$'s bias cannot lower other's assigned final scores. We show that bias also does not effect the post-regrading expected score $\max\{r_{j}^{*}(\cdot),y_{j}\}$. Thus biasing reports does not provide any competitive incentives. Her grading performance score depends only on the assigned final scores on the papers she graded, and hence it is unaffected by bias too. EPBI is independent of the condition on $w_{ik}$s.

Intuitively, two forces drive the EPRM result. 
\begin{itemize}[noitemsep,leftmargin=*,topsep=0pt,parsep=0pt,partopsep=0pt]
\item The link between $i$'s grading performance score and her marginal contribution to accurate grading plays a crucial role. 
A lower grading reliability of $i\in G(j)$ invariably lowers $i$'s marginal contribution to accurate score-assignment on paper $j$. This lowers $i$'s grading performance score and hence, her total utility. 
\item The score-assignment function and our regrading assumption (see \Cref{ass:regrading}) are crucial too. As mentioned previously, under our score-assignment function, grader $i$'s noisier grading leads to a noisier assigned grade on paper $j$. The noise moves the assigned grade above or below the true grade. Higher is the noise, larger is the potential movement in either direction. Grader $i$ determines the magnitude of the noise, but not the direction in which the noise moves the assigned grade. By selectively asking for regrades, student $j$ keeps any undeserved high grades and reverses any low grades that result from the noise. Thus, $i$'s noisier grading ends up increasing $j$'s grades post regrading-requests. Given $i$ dislikes when $j$ gets higher grades, this decreases $i$'s utility in expectation. Thus, $i$'s competitiveness also fuels $i$'s desire for an accurate grading. 
\end{itemize}

\noindent
Deriving the EPRM condition requires a bound on $w_{ik}$s. This is because the choice of reliability of grader $i$ affects the final grades of whoever she grades, and the marginal contributions (thus, grading performance scores) of her co-graders. 
We show that the condition on $w_{ik}$s is sufficient to ensure that the collective weight on other's grading performance scores never outweighs a competitive student's regard for her own performance score, irrespective of other's actions and noise. In the proof, we also show that the sufficient condition on the $w_{ik}$'s can be further weakened to a sum over only her co-graders. We kept the condition as mentioned in the theorem statement for simplicity and explainability.



%

The weight $\alpha$ that an instructor assigns in \mechabbrv\ (see Step~\ref{step:payment} of \Cref{alg:trupeqa}) on the peer-grading performance score can vary across different instructors. It is, therefore, desirable to have a score-assignment function that is robust to any choice of the weight $\alpha$ while retaining the two properties above. 
%
%
%
%

Following our discussions around \Cref{eq:r-general-k}, we show in our following result why the ERM score-assigment function is not the optimal choice from the WA class in a world where reliability needs to be incentivized. Rather \mechabbrv's score-assignment function that weighs the de-biased scores by the inverse of the square root of estimated variance, works better.

\begin{theorem}[Uniqueness]
\label{thm:grade-unique}
 Assume for every $i \in N$, $\exists j \in G^{-1}(i)$ s.t.\ $w_{ij} > 0$. A peer-grading mechanism $M = \langle G, \mathbf{r^*}, \mathbf{t} \rangle$, where $r_j^* \equiv r_j^\text{WA}, \forall j \in N$ (\Cref{eq:r-general-k})
 with any performance score function $t$ satisfies EPRM for every peer-grading performance score weight $\alpha > 0$ and for all realizations of $y_j$ and $\tilde{y}_j^{(i)}$, $i \in N, j \in P_i$, if and only if $\lambda_i = c_i / \sigma_i$, where $c_i > 0$ is any arbitrary constant independent of $\sigma_i$.
\end{theorem}
Note that, $\sqrt{\hat{\tau}_i} \propto 1/\sigma_i$.\footnote{The proportionality constants depend on the realized values of $y_j$ and $\tilde{y}_j^{(i)}$, $j \in P_i$, after division by $\sigma_i$.}
Therefore, in the class of weighted average score computing function, ISWDM score-assignment function, used by \mechabbrv\, uniquely (upto constant multipliers) ensures EPRM for flexible performance score weight $\alpha > 0$.
This result shows why our score-assignment function is special, irrespective of the choice of grading performance scores. 

At the risk of oversimplification, here is an intuition about how this result works. For the class of weighted average (WA) score-assignment functions, let us consider how the weights affect the {\em post-regrading} score.
\begin{equation}
 \max \left \{r_j^\text{WA}(\mathbf{\tilde{y}}^{G(j)}_{j}, \hat{\boldsymbol{\theta}}_{G(j)}),y_j \right \} = y_j + \max \left \{\frac{\lambda_0 (\mu-y_j)+\sum_{i \in G(j)}\lambda_i(\tilde{y}_{j}^{(i)}-\hat{b}_{i}-y_j)}{\lambda_0+\sum_{i\in G(j)}\lambda_i},0 \right \} 
\end{equation} 
The first term on the RHS is the true score $y_j$ which is independent of grader $i$'s actions. We will focus on how grader $i$'s choices affect the numerator and the denominator of the second term, which is non-negative. The $(\tilde{y}_{j}^{(i)}-\hat{b}_{i}-y_j)$ terms are approximately a measure of the noise present in the signals that grader $i$ observed for paper $j$, which has a variance of $\sigma_i^2$. 
But,  $\lambda_i = c_i / \sigma_i$ uniquely makes the product $\lambda_i(\tilde{y}_{j}^{(i)}-\hat{b}_{i}-y_j)$ independent of grader $i$'s chosen $\sigma_i$, for all values of $\sigma_i$.
This is true for all her co-graders too. 
Hence the numerator is independent of the variance of the graders, which is the first step of the proof. 

The denominator is the sum of positive numbers. The term $\lambda_i = c_i / \sigma_i$ guarantees that when $\sigma_i$ increases the whole fraction increases. Thus, noisier grading ends up increasing the post-regrading score $\max \left \{r_j^\text{WA}(\mathbf{\tilde{y}}^{G(j)}_{j}, \hat{\boldsymbol{\theta}}_{G(j)}),y_j \right \}$. In the proof, we formalize this intuition, while accounting for what information is available to grader $i$ when she contemplates how her actions affect post-regrading scores.

\section{Efficiency Under Costly Effort}
\label{sec:costlyeffort}

In this section, we assume that increasing reliability is effort-intensive. Now that reliability is costly, what is a desirable level of reliability (effort) from a social point of view? 
To calculate the social utility, we sum the grading-accuracy of all exams and subtract the total effort-cost from it. We show that a modification of the grading performance score $\mathbf{t}^*$ of \mechabbrv\ implements the socially optimal level of costly effort. 

We assume that the graders have a uniform weight for the other-regarding component in their utility ($w_{ij} = w, \forall\ i, j \in N$) and is a common knowledge.


\smallskip \noindent
\textbf{Costly effort.}
We assume that each exam has at least two questions in an exam, and the grading difficulty is dependent on the question type.\footnote{For instance, in a physics exam, a question on the general theory of relativity is more difficult to grade than that on the Newton's laws of motion.} We also assume that all graders face the same effort-cost function $c_k$ while grading the question $k$ of the exam. Hence, in every paper (answerscript) grading question $k$ will have the cost $c_k(\tau_{ik})$ for grader $i$, and we allow $c_k \ne c_{k'}$ for questions $k \ne k'$. We assume that graders can observe $c_k$ while they grade question $k$, but the mechanism designer cannot.

The grader can choose different reliabilities for different questions within a paper, but she grades any particular question at same reliability across all papers: $\tau_{ik}$ remains the same for question $k$ on all the papers $i$ grades. Hence, grader $i$ chooses a reliability vector $\tau_i = (\tau_{ik}, k \in Q)$, where $\tau_{ik}$ is the reliability specific to the question $k$ of the paper and $Q$ is the set of all questions in a paper. The estimated reliability for grader $i$ for question $k$, $\hat{\tau}_{ik}$, is computed from her performance on the $k$th question in the probe papers. Reliability is bounded above, i.e., $\tau_{ik} \in [0, \bar{\tau}], \forall i \in N, k \in Q$. We summarize our assumptions below.
\begin{enumerate}[noitemsep,topsep=0pt,parsep=0pt,partopsep=0pt]
 \item The cost $c_k: [0, \bar{\tau}] \to \mathbb{R}_{\geqslant 0}$ is  convex, increasing, and equal for all graders $i \in N$. 
\item The course instructor does not know $c_k$, only the graders do.
\end{enumerate}

\smallskip \noindent
\textbf{Social utility of grading.} 
For any question on a non-probe paper, we assume that the social planner (e.g., the instructor) cares about two things: (a)~the accuracy of the final score (measured by the reward function $R(r_j^*, y_j)$) and (b)~the total cost of grader-effort. 

We presume that {\em if the social planner was aware of the cost functions of grading}, she would have recommended a joint strategy profile $(\tau_i, \tau_{-i})$ that maximizes some linear combination of the reward and cost factors, which we call the {\em social utility}. Formally, the social utility of grading paper $j$ is written as
\begin{equation}
 \label{eq:social-utility}
 E_{y_j} E_{ (\tilde{y}_{j}^{(k)}|y_j) \sim \mathcal{F}(y_{j}+b_{k},1/\tau_{k}), k \in G(j) }  \left(  \beta R(r_j^*(\mathbf{\tilde{y}}^{G(j)}_{j}, \hat{\boldsymbol{\theta}}_{G(j)}), y_j) - \sum_{i \in G(j)} \sum_{k \in Q} c_k(\tau_{ik}) \right),
\end{equation}
where $\beta>0$ determines the relative weight between the two factors.  
The final social utility is the sum of this over all non-probe papers. The socially optimal effort for any question depends on its cost. When the cost is private information, it impossible to dictate socially optimal effort centrally.

\smallskip \noindent
\textbf{Aligning social and individual incentives.} 
There are three challenges on the way to aligning social and individual incentives. We discuss these below along with their solutions.
\begin{itemize}[noitemsep,leftmargin=*,topsep=0pt,parsep=0pt,partopsep=0pt]
\item {\em An instructor would care about the accuracy in grade-allocation, but why would peer-graders care about the same?} \mechabbrv's grading performance score solves this. It forces subjects to internalize accuracy in their decisions by paying each grader their marginal contribution to grading accuracy. 
\item Each competitive grader wants lower scores for others as part of her other-regarding utility. Clearly this is not aligned with the social utility of grading and becomes a source of externality. We solve this by suggesting a modified grading performance score below, that additionally compensates graders for any potential losses from their other-regarding utility component.
\item The solution to the point above presents a new challenge. The other-regarding utility component would be different for each grader $i$, as their reference groups $N \setminus \{i\}$ would naturally be different. Thus they will be compensated different amounts. {\em Would this change the ordinal ranking of students in the class from that of \mechabbrv?} We show that the answer is {\em no} (\Cref{lemma:order-inv}).
\end{itemize}

\smallskip \noindent
\textbf{Modified grading performance score.} Let the post-regrading request score be $g_i=\max\{r_i,y_i\}$. We propose the modified grading performance score 
\begin{equation}
 \label{eq:mod-grading-score}
 \pi_i := t_i + w \sum_{j \in N \setminus \{i\}} (g_j + \pi_j),
\end{equation}
where $t_i$ is the original \mechabbrv\ grading performance score. The additional terms on the RHS compensate for the other-regarding component in grader $i$'s utility. 
Though this simplifies the net utility of grader $i$, the simplicity comes at a price: if $i$ and $j$ are co-graders then $\pi_i$ has been described as a function of $\pi_j$ and vice-versa! How is the designer supposed to decide the values of $\pi_i$ and $\pi_j$ given the interdependency? We show that $\pi_i$ has an alternative expression that is independent of $\pi_j$s. 
\begin{align}
 \pi_i &= \frac{t_i + w \sum_{j \in N \setminus \{i\}} g_j + w \pi}{1+w}, \label{eq:mod-performance score}\\
 \text{where, } \pi &= \frac{t + w (n-1) g}{1 - w (n-1)}, \text{ and } t = \sum_{i \in N} t_i, \ g = \sum_{i \in N} g_i, \ w \neq \frac{1}{n-1}. \label{eq:sum-performance score}
\end{align}

\smallskip \noindent
\textbf{The game of peer-grading.} 
The modified \mechabbrv\ mechanism induces a game among the peer-graders after all the answerscripts of the exam have been submitted. The players (the graders) choose their reliabilities as their strategies to maximize their utility. 
Grader $i$'s utility is given by
\begin{equation}
 \label{eq:utility-with-costlyeffort}
 u_i = g_i + \pi_i - w \sum_{j \in N \setminus \{i\}} (g_j + \pi_j) - \sum_{j \in G^{-1}(i)}  \sum_{k \in Q} c_k(\tau_{ik}) = g_i + t_i - \left|G^{-1}(i)\right|  \sum_{k \in Q} c_k(\tau_{ik}),
\end{equation}
which is common knowledge. Players simultaneously choose their reliability vectors $\tau_i$. The score assignment and performance score functions, that map players' strategies to players' utilities, are also common knowledge.
The following result shows that it retains the same order of the scores as in the \mechabbrv\ performance score.
\begin{lemma}[Order Invariance]
 \label{lemma:order-inv}
Fix a profile of player strategies and true scores in the peer-grading game. The modified \mechabbrv\ performance score $\pi$ retains the same order among the students as the original \mechabbrv\ performance score $t$.
\end{lemma}
\begin{proof}
 $g_i + \pi_i > g_k + \pi_k \Leftrightarrow g_i + \frac{t_i + w \sum_{j \in N \setminus \{i\}} g_j + w \pi}{1+w} > g_k + \frac{t_k + w \sum_{j \in N \setminus \{k\}} g_k + w \pi}{1+w} \Leftrightarrow \frac{g_i + t_i + w(g+\pi)}{1+w} > \frac{g_k + t_k + w(g+\pi)}{1+w} \Leftrightarrow g_i + t_i > g_k + t_k$.
\end{proof}

The following result shows that the modified grading performance score $\mathbf{\pi} := (\pi_i, i \in N)$ implements the social optimal level of effort for every paper $j$ in a pure Nash equilibrium. The proportion of non-probe papers (which is fixed once the mechanism is announced) is denoted by $p_\text{NP}$.

\begin{theorem}
\label{thm:costly}
For every paper $j \in N$, if the course designer uses the modified grading score $\pi_i$ and sets $ \alpha = \frac{\beta}{p_\text{NP}} $, every maxima of the expected social utility is a Pure Strategy Nash Equilibrium (PSNE) of the induced game among the graders of the paper.
\end{theorem}

\begin{proof}
The modified \mechabbrv\ already compensates for the other-regarding component and makes it inconsequential.
The residual performance score $t_i$ is the sum of performance scores $t_i^j$ from each paper $j \in G^{-1}(i)$. Now, $t_i^j=\alpha(W_j^{*}-W_j^{(-i)*})$, and $W_j^{(-i)*}$ does not depend on $i$'s reliability $\tau_i$. Hence the part of the  $i$'s utility expression that depends on $\tau_i$ and is related to grading paper $j$ is (using the shorthand $c(\tau_i) \equiv  \sum_{k \in Q} c_k(\tau_{ik})$:

\begin{equation}
\alpha W_j^* -  \sum_{k \in Q} c_k(\tau_{ik}) = \alpha R(r_j^*(\mathbf{\tilde{y}}^{G(j)}_{j}, \hat{\boldsymbol{\theta}}_{G(j)}), y_j) - c(\tau_i),
\end{equation}
where
$\mathbf{\tilde{y}}^{G(j)}_{j}$ is the profile of all scores given by $G(j)$. Thus it depends on the bias and reliability of co-graders, which is chosen strategically and simultaneously. 
Grader $i$ is uncertain whether paper $j$ is a probe versus a non-probe paper, and only the latter provides a performance score. Since the proportion $p_\text{NP}$ of non-probes is announced by the mechanism, $i$ assigns a probability $p_\text{NP}$ to any paper being a non-probe. For any choice of bias and reliability by all the graders on paper $j$, the expected reward to the mechanism designer is denoted by $ E_{y_j} E_{ (\tilde{y}_{j}^{(k)}|y_j) \sim \mathcal{F}(y_{j}+b_{k},1/\tau_{k}), k \in G(j) } R(r_j^*(\mathbf{\tilde{y}}^{G(j)}_{j}, \hat{\boldsymbol{\theta}}_{G(j)}), y_j)$.\footnote{The distributions of these random variables are common knowledge of the graders.} From the analysis of \Cref{thm:bi-rm}, we know that this function is independent of bias under \mechabbrv. Therefore, for simplicity, we assume that every grader strategizes only on her reliability. To emphasize the strategic and simultaneous choice of reliability, we rewrite $R(\cdot)$ using the shorthand $\bar{R}(\tau_i, \tau_{-i})$.

Hence, for any reliability profile chosen by the set of graders on paper $j$, the part of the expected utility of grader $i$ that depends on $\tau_i$ is
\begin{equation}
 \label{eq:ind-utility}
 U^j_i(\tau_i, \tau_{-i}) =  \alpha \cdot p_\text{NP} \cdot \bar{R}(\tau_i, \tau_{-i}) - c(\tau_i).
\end{equation}

When $\alpha = \frac{\beta}{p_\text{NP}}$, grader $i$'s relevant utility becomes 
 $U^j_i(\tau_i, \tau_{-i}) = \beta \bar{R}(\tau_i, \tau_{-i}) - c(\tau_i)$.
%
Let the social optimal be obtained at $\tau^*= (\tau_k^*, k \in G(j))$.
Then, it must be that $\beta \bar{R}(\tau_{i}^*, \tau_{-i}^*) - c(\tau_i^*) \geqslant \beta \bar{R}(\tau_{i}, \tau_{-i}^*) - c(\tau_{i}) $ for all reliabilities $\tau_{i}$ of player $i \in G(j)$, because, if any grader $i$ could increase her expected utility by choosing any other $\tau_i$, then it is easy to show that it will contradict the optimality of the social utility at $\tau^*$. Thus, clearly if all except $i$ choose $\tau_{-i}^*$, player $i$ cannot do any better than choosing $\tau_{i}^*$. Thus, $\tau^*=(\tau_k^*, k \in G(j))$ is a PSNE of this game.
\end{proof}
As mentioned in the proof above, the property of EPBI is retained even in this setting with costly efforts since the change in the utility due to cost is independent of bias.
In the next section, we present our experimental study that tests some of our hypotheses made in the earlier results and verifies its practical usability.

\section{Experimental study}
\label{sec:human}

One major objective of the experimental study was to understand the practical trade-off between the theoretical desirability of \mechabbrv\ against the simple and widely used peer-grading mechanism. As a comparison candidate, we chose the {\em median mechanism}, where the score-assignment of a paper is done based on the median of the given scores of the peer-graders evaluating that paper. This mechanism is used in practice for peer-grading MOOCs, e.g., Coursera (\url{https://www.coursera.org}) uses this mechanism across multiple courses~\citep{CourseraaccessedFeb2021}. 

Another objective of this study was to test two of our modeling assumptions: if bias and reliability were indeed identical in probes and non-probes, and, if competitive ($w_{ij} \geqslant 0$) preferences are a good model of the peer-grader behavior. 

Finally, the theoretical desirability of \mechabbrv\ is established on restrictive assumptions about the domain of true and given scores, player utilities, and strategies. These assumptions approximate reality instead of describing it. How well does \mechabbrv\ perform in a real-life exercise, where the scores and signals come from a bounded interval, or when player's utilities are competitive but not necessarily linear?


\subsection{Experimental Design} 
\label{sec:expt-design}

We ran two experimental sessions: one with the median scoring mechanism ($27$ students),  another with the \mechabbrv\ mechanism ($42$ students). We recruited subjects through two open-calls to undergraduate students enrolled in a computing course (Prog101). The open calls did not contain any particulars of the two sessions. Every student who signed up for participation was assigned to one of the two sessions.

The experimental environment is not an exact replication of model assumptions, rather a replication of how a real-life peer-grading scenario would look like. In many classes, instructors grade on a curve: final numerical scores are converted to letter-grades (A to D) based on relative rankings. Grading on a curve creates a competitive classroom-environment that we wanted to replicate. We told participants that their total score is the sum of their peer-evaluated score and grading performance score. We paid subjects by the relative ranking of their total scores in the class, in both the sessions. The students who ranked in the first quartile of the total scores received M~650,\footnote{M = Indian Rupee (\rupee), a difference of M 200 is significant for a student.}
the next three quartiles received M~450, M~250, and M~50 respectively. They also received a show-up fee of M~50, irrespective of their total score. The monetary payments were placeholders for grades A to D in a class that grades on a curve: high relative performance resulted in high rewards.\footnote{The ethics committee did not allow us to use university grades in the Prog101 class as incentives.}

In the median mechanism session, the grading performance score of all subjects were set to zero. The total-score ranking was identical to their peer-evaluated-score ranking. Thus, a subject could decrease others' scores on the peer-evaluation task to increase her relative ranking and payment. 

The \mechabbrv\ session used the \mechabbrv\ assignment and grading performance scores. Thus, manipulation on the peer-evaluation task risked getting a lower performance and total score, which would result in a lower payment.

The instructions and incentive-scheme, included in \Cref{sec:instructions}, were explained in detail before each of the sessions began. In both sessions, we used numerical examples in our explanation. For \mechabbrv, we showed the relation between performance score and grading reliability through a graph and verbally summarized the monotonic relationship. 

We conducted both sessions during the weekly Prog101 labs, that happen in a large computer lab. Our study lies at the intersection of {\em Lab} and {\em Field} experiments. We are interested in peer-grading behavior and the students are our population of interest. In this study, we observed our population of interest in their {\em naturally occurring environments}, like in Field experiments.



In both sessions, we asked subjects to peer-grade the same weekly class-quiz. We partitioned each quiz into three sub-quizzes (by treating one(two) question(s) of the quiz as a sub-quiz\footnote{The quiz had more than three questions.}), and divided each session into {\em three} rounds.  In every round, the subjects were asked to peer-grade five sub-quizzes (each corresponding to one of five of her anonymous peers). At the end of each round, subjects saw: (a)~how peers had evaluated her performance on the sub-quiz, (b)~her assigned score (median-scoring or \mechabbrv), and (c)~how her co-graders that round had evaluated the sub-quiz.

Within every sub-quiz, some (and not all) of the questions were `regradable'. The students could raise a regrading request for only those questions at the end of the session. In the \mechabbrv\ sessions, only the regradable questions were incentivized by the grading performance score. The non-regradable questions used the same assignment function but did not have any grading performance score.

We also graded all the papers ourselves (the instructor graded all of them), and we considered these scores to be the {\em true scores}. The difference between mechanism assigned scores and true scores is a measure of the quality of these mechanisms.

\subsection{Hypotheses and results}
\label{sec:hyp-results}

Bias is the difference between the true score and the peer-assigned score. It measures the average direction and magnitude of manipulation. \mechabbrv\ assumes that bias is zero or positive: subjects generally do not manipulate scores upwards (i.e., do not collude). Our first hypothesis builds on this assumption.
\begin{hypothesis}
\label{hyp:bias-sign}
Score-manipulation is not collusive.
\end{hypothesis}

Our second hypothesis suggests that bias should be higher in the last round for two reasons. First, most repeated interactions have an end-game effect: selfish behavior unravels when no future interactions remain. Second, subjects who have experienced score-manipulation by others might retaliate as a punishment or reciprocal strategy in the later rounds of the treatment.
\begin{hypothesis}
\label{hyp:manipulation-later}
Score-manipulation or bias peaks in the last round of the game.
\end{hypothesis}

In \Cref{tab:Results-from-Median,tab:Results-from-TRUEQA}, we summarize the bias in individual grading behavior in the three rounds of both treatments. To compare across questions and rounds, we normalize bias by the total score of the corresponding question.

\begin{table}[H]
\begin{centering}
{\footnotesize{}}%
\begin{tabular}{|c|c|c|c|c|c|}
\hline 
 &  &  & {\footnotesize{}Total} & \multicolumn{2}{c|}{{\footnotesize{}Avg bias (\% of Total grade)}}\tabularnewline
\cline{5-6} 
 & {\footnotesize{}Subjects} & {\footnotesize{}Round} & {\footnotesize{}grade} & {\footnotesize{}regradable} & {\footnotesize{}non-regradable}\tabularnewline
\hline 
\hline 
\multirow{6}{*}{Median} & \multirow{6}{*}{{\footnotesize{}27}} & {\footnotesize{}1} & {\footnotesize{}1+1} & {\footnotesize{}-0.4\%} & {\footnotesize{}-0.6\%}\tabularnewline
\cline{3-6} 
 &  &  &  & {\footnotesize{}(-4\%,3.2\%)} & {\footnotesize{}(-3.1\%,1.9\%)}\tabularnewline
\cline{3-6} 
 &  & {\footnotesize{}2} & {\footnotesize{}2+2} & {\footnotesize{}1.7\%} & {\footnotesize{}1.2\%}\tabularnewline
\cline{3-6} 
 &  &  &  & {\footnotesize{}(-1\%,4.4\%))} & {\footnotesize{}(-0.8\%,3.1\%)}\tabularnewline
\cline{3-6} 
 &  & {\footnotesize{}3} & {\footnotesize{}2+2} & {\footnotesize{}16.6\%} & {\footnotesize{}16.4\%}\tabularnewline
\cline{3-6} 
 &  &  &  & {\footnotesize{}(12\%,21.2\%)} & {\footnotesize{}(12\%,20.7\%)}\tabularnewline
\hline 
\end{tabular}{\footnotesize\par}
\par\end{centering}
{\caption{Average bias from 3 rounds grading under the
Median mechanism. We report the 95\% confidence intervals below the averages.}
\label{tab:Results-from-Median}
}
\end{table}
%
%
%
In each round, every subject graded a regradable and a non-regradable question. The average bias is statistically identical to zero for the first two rounds, and significantly positive in the third round. This is true for both the regradable and non-regradable questions. Thus, the bias is either zero, or positive, and we cannot reject \Cref{hyp:bias-sign}. The average bias is also significantly higher in the third round, confirming \Cref{hyp:manipulation-later}. This holds for both regradable and non-regradable questions. 
%
%
\begin{table}[H]
\begin{centering}
\begin{tabular}{|c|c|c|c|c|c|}
\hline 
 &  &  & {\footnotesize{}Total} & \multicolumn{2}{c|}{{\footnotesize{}Avg bias (\% of Total grade)}}\tabularnewline
\cline{5-6} 
 & {\footnotesize{}Subjects} & {\footnotesize{}Round} & {\footnotesize{}grade} & {\footnotesize{}regradable} & {\footnotesize{}non-regradable}\tabularnewline
\hline 
\hline 
\multirow{6}{*}{\mechabbrv} & \multirow{6}{*}{{\footnotesize{}42}} & {\footnotesize{}1} & {\footnotesize{}1+1} & {\footnotesize{}-0.6\%} & {\footnotesize{}-0.5\%}\tabularnewline
\cline{3-6} 
 &  &  &  & {\footnotesize{}(-2\%,1\%)} & {\footnotesize{}(-1.9\%,0.9\%)}\tabularnewline
\cline{3-6} 
 &  & {\footnotesize{}2} & {\footnotesize{}2+2} & {\footnotesize{}0.6\%} & {\footnotesize{}-0.3\%}\tabularnewline
\cline{3-6} 
 &  &  &  & {\footnotesize{}(0\%,1.2\%)} & {\footnotesize{}(-1.2\%,0.6\%)}\tabularnewline
\cline{3-6} 
 &  & {\footnotesize{}3} & {\footnotesize{}2+2} & {\footnotesize{}15.8\%} & {\footnotesize{}15\%}\tabularnewline
\cline{3-6} 
 &  &  &  & {\footnotesize{}(12.3\%,19.2\%)} & {\footnotesize{}(11.2\%,18\%)}\tabularnewline
\hline 
\end{tabular}{\footnotesize\par}
\par\end{centering}
{\caption{Average bias from 3 rounds grading under the
\mechabbrv\ mechanism. We report the 95\% confidence intervals under the
averages.}
\label{tab:Results-from-TRUEQA}
}
\end{table}

The low bias in grading in the first two rounds parallels the results on honest reporting from the ``die-roll in person and report'' studies \citep{mazar2008dishonesty,fischbacher2013lies}. In these studies, subjects roll a die privately, self-report the outcome, and get paid based on the report. \citet{fischbacher2013lies} report that only 20\% of people lie to the fullest extent, 39\% choose to be honest, and a sizable proportion cheats only marginally. Lying aversion~\citep{dufwenberg2018lies}, caring about lie-credibility, and a notion of self-concept maintenance~\citep{mazar2008dishonesty} are potential reasons for why people do not lie completely even under full anonymity.

How do the two mechanisms perform? The median assignment rule, due to its robustness to outliers, is immune to insincere grading as long as only a minority of graders are insincere. \mechabbrv\ is bias invariant (EPBI), incentivizes effort, and should outperform the Median mechanism. We use the accuracy of the mechanism-assigned scores as a metric of relative performance. Given subjects graded most insincerely in the third round, we use this round to test \Cref{hyp:better-final-score}. 
\begin{hypothesis}
\label{hyp:better-final-score}
In the presence of strategic manipulation, the final score assigned under \mechabbrv\ should be closer to the true scores, than that assigned under median-scoring.
\end{hypothesis}

 In \Cref{tab:Comparison-between-Median-TRUPEQA}, we present the means of fractional-difference and squared fractional-difference between the mechanism-assigned score and true score.
 Thus, for the former we calculate $d_j=(\text{true score}_j - \text{mechanism assigned score}_j)/ \text{total score}_j$ on student $j$'s third-round sub-quiz, and then take the average over all $j$. Similarly, the latter is the average of $d_j^2$. We find the true score on an exam by grading it ourselves.

\begin{wraptable}{r}{0.5\linewidth}
 \begin{center}
 \vspace{-20pt}
\begin{tabular}{|c|c|c|}
\hline 
\multirow{2}{*}{} & \multirow{2}{*}{{Median}} & \multirow{2}{*}{{\mechabbrv}}\tabularnewline
 &  & \tabularnewline
\hline 
\hline 
{Mean Difference} & {14.8\%} & {-1.2\%}\tabularnewline
\cline{2-3} 
{Mean Squared Difference} & {14.8\%} & {0.6\%}\tabularnewline
\cline{2-3} 
{$N$} & {27} & {42}\tabularnewline
\hline 
\end{tabular}
\caption{Difference and squared-difference.}
\label{tab:Comparison-between-Median-TRUPEQA}
\vspace{-20pt}
\end{center}
\end{wraptable}

The average difference between true and mechanism assigned scores is 14.8\% under Median and only -1.2\% under \mechabbrv. The negative sign indicates that \mechabbrv\ assigned slightly higher scores than the true score. Both difference and squared difference are significantly smaller under \mechabbrv. Under the median mechanism, the difference and squared difference were equal because $d_j$ almost always took values of 0 or 1.

%
The median mechanism assigned lower than true grade (assigned a 1 instead of a 2) for 15\% (4 out of 27) of the sub-quizzes. In comparison, the \mechabbrv\ mechanism was (almost) always point-precise: only one sub-quiz (out of 41) assigned a grade of 0.5 points higher. The number of regrading requests in the median and \mechabbrv\ sessions were 4/27 and 3/41 respectively, a difference that is statistically significant.

One of the crucial assumptions of \mechabbrv\ was that bias and noise are invariant across probes and non-probes.

\begin{hypothesis}
\label{hyp:bias-equal}
Bias and noise are identical in probes and non-probes.
\end{hypothesis}

We tested if bias was different across probe and non-probe questions from the \mechabbrv\ session. We pooled across three rounds to maximize power. Our statistical tests fail to reject \Cref{hyp:bias-equal}.  In a t-test, we could not reject the equality of average bias across regradable probe and non-probe questions: the p-value was 0.19. The p-value was 0.16 when we ran the same test for the non-regradable questions.

We also tested if the mean squared deviation (noise) was different across probe and non-probe questions from the \mechabbrv\ session. We could not reject the equality of noise across probe and non-probe questions. The corresponding p-values were 0.86 and 0.31 respectively for the regradable and non-regradable questions. 

\section{Conclusion}
\label{sec:concl}

We introduce a new mechanism, \mechabbrv, that uses a score-assignment rule and grading performance scores to incentivize competitive graders. The rule and the performance score guarantee unbiased grades. They also guarantee that any grader's utility increases monotonically with her grading reliability, irrespective of her competitiveness and how her co-graders act. Our assignment rule is unique in its class to satisfy this utility-reliability monotonicity while allowing flexibility in how large performance scores need to be. When grading is costly, a special version of \mechabbrv\ implements the socially optimal effort-choices in an equilibrium of the peer-evaluation game among co-graders. Finally, in our classroom experiments, \mechabbrv\ outperforms the popular median mechanism.

\section*{Acknowledgments}
 This work has been supported by the Indian Institute of Technology Kanpur under the grant number 2017198. We would like to thank the Institutional Ethics Committee (IEC) of the Indian Institute of Technology Kanpur for providing us with the opportunity to run the human subject study with the students of the institute via the IEC Communication Number: IITK/IEC/2020-21/II/30. 
 We also thank Bikramaditya Datta, Debasis Mishra, and the seminar/ workshop participants at UC Davis, Academia Sinica, and WED 2019 (ISI Delhi) for their many valuable comments.

\bibliographystyle{ACM-Reference-Format}
\bibliography{abb,ultimate,swaprava}

\pagebreak

\appendix
\section*{Appendices}

\section{Omitted Proofs}
\label{sec:proofs}

\subsection{Proof of \Cref{thm:bi-rm}}
\label{sec:proof-bi-rm}
By \Cref{ass:regrading}, the student knows her $y_j$ perfectly and if $r_j^* \geqslant y_j$, she does not raise a regrading request. \mechabbrv\ will assume $r_j^*$ to be the true score and design the peer-grading performance score accordingly when there is no regrading request. The student asks for regrading {\em only if} $r_j^* < y_j$.
The utility of grader $i$ after the regrading requests have been addressed is (we have omitted the arguments of the functions in \Cref{eq:utility} where it is understood) therefore
\begin{align}
\label{eq:utility-2}
u_{i}(\cdot) &= \max\{r_{i}^{*}(\cdot),y_{i}\}+t_{i}- \left ( \sum_{j \in G^{-1}(i)}w_{ij} \max\{r_{j}^{*}(\cdot),y_{j}\}+ \sum_{k \in CG_i \setminus \{i\}} w_{ik} t_k \right) - \phi(\cdot)
\end{align}
We decomposed the utility expression to gather together the terms that are affected by the choices of $b_i, \tau_i$ of student $i$. They are (a)~the exam scores of the papers graded by $i$ (first term in the parentheses), and (b)~the peer-grading performance score of the co-graders of $i$ (second term in the parentheses). The function $\phi$ is the remaining part of $u_i$ that is independent of $b_i, \tau_i$.

We prove that \mechabbrv\ is EPBI and EPRM in {\em four} steps. 
First, we observe that the first term on the RHS is independent of the values of $b_i$ and $\tau_i$. In the second step, we show that each summand $\max\{r_{j}^{*}(\cdot),y_{j}\}$ in the first summation term is independent of $b_i$ and decreasing in $\tau_i$. The third step shows that $t_{i}$ is independent of $b_i$ and increasing in $\tau_i$, and the fourth step shows that this conclusion is true even for $t_{i} - \sum_{k \in CG_i \setminus \{i\}} w_{ik} t_k$ for the sufficient condition of the theorem. 

\paragraph{Step 1: $\max\{r_{i}^{*}(\cdot),y_{i}\}$ is independent of the values of $b_i$ and $\tau_i$.}
This is obvious since student $i$ does not grade her own paper and hence she has no control on the grade given by \mechabbrv\ on her paper.

\paragraph{Step 2: The expected value of $\max\{r_{j}^{*}(\cdot),y_{j}\}$ is independent of $b_{i}$ and increasing in $\sigma_{i}$.}
Recall that the score-assignment function for \mechabbrv\ is ISWDM (\Cref{def:score-welfare})
\[r_{j}^{*}(\mathbf{\tilde{y}}^{G(j)}_{j}, \hat{\boldsymbol{\theta}}_{G(j)}) = \frac{\sqrt{\gamma}\mu+\sum_{i \in G(j)}\sqrt{\hat{\tau}_{i}}(\tilde{y}_{j}^{(i)}-\hat{b}_{i})}{\sqrt{\gamma}+\sum_{i\in G(j)}\sqrt{\hat{\tau}_{i}}}.\]
The final grade after regrading is 
\begin{equation}
 \label{eq:r-diff}
 \max\{r_{j}^{*}(\cdot),y_{j}\}=\max\{r_{j}^{*}(\cdot)-y_{j},0\}+y_{j}.
\end{equation}
Grader $i$'s estimated bias is given by $\hat{b}_i = \frac{\sum_{k\in P_i}(\tilde{y}_j^{(i)}-y_j)}{x}$, where $x = |P_i|$. In \mechabbrv, we use the same number $K/2$ as $|P_i|$, for all $i$. Hence, $x = K/2$, is a constant in our analysis.

Given our model of peer-reports, $\tilde{y}_j^{(i)}=y_j+b_i+n_{ij}$, where $n_{ij} \sim \mathcal{F}(0,1/\tau_{i})$ is a noise term. Hence, it is easy to show that $\hat{b}_i = b_{i}+\frac{\sum_{k\in P_i} n_{ik}}{x}$ and $\frac{1}{\hat{\tau}_i}=\hat{\sigma}_{i}^{2} = \frac{\sum_{k \in P_i}(n_{ik}-\frac{1}{x}\sum n_{ik})^{2}}{x}$, where $n_{ik} \sim \mathcal{F}(0,\sigma_{i}^2)$.

Substituting these values we get the expression for
\[r_{j}^{*}(\cdot)-y_{j} = \frac{\sqrt{\gamma}(\mu-y_{j})+\sum_{l\in G(j)}\sqrt{\hat{\tau}_{l}}(n_{lj}-\frac{\sum_{k\in P_l}n_{lk}}{x})}{\sqrt{\gamma}+\sum_{l\in G(j)}\sqrt{\hat{\tau}_{l}}}.\]
Note that $z_{j}=\sqrt{\gamma}(\mu-y_{j})$ is a $\mathcal{F}(0,1)$ variable, that is independent of all the other variables in the expression. In the following, we take the expectation of the term $\max\{r_{j}^{*}(\cdot)-y_{j},0\}$ w.r.t.\ $z_j$ and show that it is independent of $b_i$ and increasing in $\sigma_i = 1/\sqrt{\tau_i}$, which implies that irrespective of the values of the other graders' biases and reliabilities, it is best for grader $i$ to reduce her $\sigma_i$ to increase this component of her utility (since the term comes with a negative sign in the utility expression).
\begin{align}
\lefteqn{ I_{j} =E_{z_j} \max\{r_{j}^{*}(\cdot)-y_{j},0\} } \nonumber \\
&= \int_{-\sum_{l \in G(j)}\sqrt{\hat{\tau}_{l}}(n_{lj}-\frac{\sum_{k\in P_l}n_{lk}}{x})}^{\infty}\frac{z_{j}+\sum_{l\in G(j)}\sqrt{\hat{\tau}_{l}}(n_{lj}-\frac{\sum_{k\in P_l}n_{lk}}{x})}{\sqrt{\gamma}+\sum_{l\in G(j)}\sqrt{\hat{\tau}_{l}}} f(z_{j})dz_{j} + 0 \nonumber \\
 &= \frac{1}{\sqrt{\gamma}+\sum_{l \in G(j)}\sqrt{\hat{\tau}_{l}}}\int_{-\sum_{l\in G(j)}\sqrt{\hat{\tau}_{l}}(n_{lj}-\frac{\sum_{k\in P_l}n_{lk}}{x})}^{\infty}(z_{j}+\sum_{l\in G(j)}\sqrt{\hat{\tau}_{l}}(n_{lj}-\frac{\sum_{k\in P_l}n_{lk}}{x})) f(z_{j})dz_{j}\nonumber \\
 &= \frac{1}{\sqrt{\gamma}+\sum_{l \in G(j)}\sqrt{\hat{\tau}_{l}}}\int_{0}^{\infty}v_{j}f(v_{j}-\sum_{l\in G(j)}\sqrt{\hat{\tau}_{l}} (n_{lj}-\frac{\sum_{k\in P_l}n_{lk}}{x})) dv_{j}\nonumber \\
 &= \frac{1}{\sqrt{\gamma}+\sum_{l \in G(j)}\sqrt{\hat{\tau}_{l}}}\int_{0}^{\infty}v_{j}f(v_{j}-\sum_{l\in G(j)\setminus\{i\}}\sqrt{\hat{\tau}_{l}}(n_{lj}-\frac{\sum_{k\in P_l}n_{lk}}{x})-\sqrt{\hat{\tau}_{i}}(n_{ij}-\frac{\sum_{k\in P_i}n_{ik}}{x})) dv_{j}\nonumber \\
 &= \frac{1}{\sqrt{\gamma}+\sum_{l \in G(j)}\sqrt{\hat{\tau}_{l}}}\int_{0}^{\infty}v_{j}f(v_{j}-\sum_{l\in G(j)\setminus\{i\}}\sqrt{\hat{\tau}_{l}}(n_{lj}-\frac{\sum_{k\in P_l}n_{lk}}{x})-\frac{(m_{ij}-\frac{\sum_{k\in P_i}m_{ik}}{x})\sigma_{i}}{\sqrt{\frac{\sum_{k\in P_i}(m_{ik}-\frac{1}{x}\sum m_{ik})^{2}}{x}\sigma_{i}^{2}}})dv_{j}\nonumber \\
 &= \frac{1}{\sqrt{\gamma}+\sum_{l\in G(j)\setminus\{i\}}\sqrt{\hat{\tau_{l}}}+1 / \sqrt{\frac{\sum_{k\in P_i}(m_{ik}-\frac{1}{x}\sum m_{ik})^{2}}{x}}\sigma_{i}} \times \nonumber \\
 & \qquad \int_{0}^{\infty}v_{j}f(v_{j}-\sum_{l\in G(j)\setminus\{i\}}\sqrt{\hat{\tau}_{l}}(n_{lj}-\frac{\sum_{k\in P_l}n_{lk}}{x})-\frac{(m_{ij}-\frac{\sum_{k\in P_i}m_{ik}}{x})}{\sqrt{\frac{\sum_{k\in P_i}(m_{ik}-\frac{1}{x}\sum m_{ik})^{2}}{x}}}) dv_{j} \label{expr:step2}
\end{align}
In the third equality, we have substituted $v_{j}=z_{j}+\sum_{l\in G(j)}\sqrt{\hat{\tau}_{l}}(n_{lj}-\frac{\sum_{k\in P_l}n_{lk}}{x})$, and in the fifth equality, we substituted $n_{ik} = m_{ik} \cdot \sigma_i$. Since $n_{ik} \sim \mathcal{F}(0,\sigma_{i}^2)$, we get $m_{ik}\sim \mathcal{F}(0,1)$. Note that $f$ is the density of a $\mathcal{F}(0,1)$ random variable. Hence the whole expression within the integral is independent of $\sigma_{i}$. It is easy to see that the pre-multiplied term is increasing in $\sigma_{i}$.
Hence, we conclude that the integral $I_j$ is independent of $b_i$ and increasing in $\sigma_i = 1/\sqrt{\tau_i}$. 

\paragraph{Step 3: The expected value of $t_{i}^j$ is independent of $b_{i}$ and decreasing in $\sigma_{i}$.}
We assumed in \Cref{sec:mechanisms} that the reward function is decreasing in the difference $|r_j^*-y_j|$ and the mechanism assigns reward to be zero when $r_j^* > y_j$. Hence, we calculate the condition on $y_j$ when the reward is non-zero.
\begin{align*}
 & r_{j}^{*}(\mathbf{\tilde{y}}^{G(j)}_{j}, \hat{\boldsymbol{\theta}}_{G(j)}) \leqslant y_{j} 
\iff \frac{\sqrt{\gamma}\mu+\sum_{l\in G(j)}\sqrt{\hat{\tau}_{l}}(\tilde{y}_{j}^{(l)}-\hat{b}_{l})}{\sqrt{\gamma}+\sum_{l\in G(j)}\sqrt{\hat{\tau}_{l}}} \leqslant y_{j}\\
& \iff \sqrt{\gamma}\mu+\sum_{l\in G(j)}\sqrt{\hat{\tau}_{l}}(\tilde{y}_{j}^{(l)}-\hat{{b_{l}}}) \leqslant y_{j}(\sqrt{\gamma}+\sum_{l\in G(j)}\sqrt{\hat{\tau}_{l}})\\
& \iff y_{j}\sqrt{\gamma} \geqslant \sqrt{\gamma}\mu+\sum_{l\in G(j)}\sqrt{\hat{\tau}_{l}}(\tilde{y}_{j}^{(l)}-\hat{b}_{l}-y_{j}) = \sqrt{\gamma}\mu+\sum_{l\in G(j)}\sqrt{\hat{\tau}_{l}}(n_{lj}-\frac{\sum_{k\in P_l}n_{lk}}{x})\\
& \iff y_{j} \geqslant \frac{\sqrt{\gamma}\mu
+ Z
+\sqrt{\hat{\tau}_{i}}(n_{ij}-\frac{\sum_{k\in P_i}n_{ik}}{x})}{\sqrt{\gamma}}, \quad \text{ where } Z = \sum_{l\in G(j) \setminus \{i\}}\sqrt{\hat{\tau}_{l}}(n_{lj}-\frac{\sum_{k\in P_l}n_{lk}}{x})\\
 & \qquad\qquad = \frac{\sqrt{\gamma}\mu+Z}{\sqrt{\gamma}}+\frac{(m_{i}-\frac{\sum_{k\in P_i}m_{ik}}{x})\sigma_{i}}{\sqrt{\gamma}\sqrt{\frac{\sum_{k\in P_i}(m_{ik}-\frac{1}{x}\sum m_{ik})^{2}}{x}\sigma_{i}^{2}}}
\end{align*}
Note that the RHS is independent of $\sigma_i$. Hence the limits of the integral where the reward $R$ is non-zero is also independent of $\sigma_i$.

By definition, the $W_j^{(-i)*}$ component of the performance score is independent of bias and reliability of grader $i$. Hence, we only consider the first component which is dependent on the bias and reliability of grader $i$. We will consider the integral only w.r.t.\ $y_j$ to compute $t_i^j$ and we just showed that the limits of this integral is independent of 
$\sigma_i$. Hence, if we show that the reward function $R(r_j^*,y_j)$ is independent of $b_{i}$ and decreasing in $\sigma_{i}$, then we are done. Consider the argument of the reward function
\begin{align}
 \lefteqn{ r_j^*-y_j = \frac{\sqrt{\gamma}(\mu-y_{j})+\sum_{l\in G(j)}\sqrt{\hat{\tau}_{l}}(n_{lj}-\frac{\sum_{k\in P_l}n_{lk}}{x})}{\sqrt{\gamma}+\sum_{l\in G(j)}\sqrt{\hat{\tau}_{l}}} } \nonumber \\
 &= \frac{[\sqrt{\gamma}(\mu-y_{j})+\sum_{l \in G(j)\setminus \{i\}}\sqrt{\hat{\tau}_{l}}(\tilde{y}_{j}^{(l)}-\hat{b}_{l}-y_{j})]\sqrt{\frac{\sum_{k\in P_i}(n_{ik}-\frac{1}{x}\sum_{l \in P_i} n_{il})^{2}}{x}}
 +
 (n_{j}-\frac{1}{x}\sum_{l \in P_i} n_{il})}
 {(\sqrt{\gamma}+\sum_{l \in G(j)\setminus \{i\}}\sqrt{\hat{\tau}_{l}})\sqrt{\frac{\sum_{k\in P_i}(n_{ik}-\frac{1}{x}\sum_{l \in P_i} n_{il})^{2}}{x}}+1} \nonumber \\
 &= \frac{Z_{-i} \sqrt{\frac{\sum_{k\in P_i}(m_{ik}-\frac{1}{x}\sum_{l \in P_i} m_{il})^{2}}{x}} \cdot \sigma_i
 +
  \sigma_i \cdot (m_{j}-\frac{1}{x}\sum_{l \in P_i} m_{il})}
 {X_{-i} \sqrt{\frac{\sum_{k\in P_i}(m_{ik}-\frac{1}{x}\sum_{l \in P_i} m_{il})^{2}}{x}} \cdot \sigma_i +1} \label{expr:step3}
\end{align}
In the last equality, we substituted $Z_{-i} = [\sqrt{\gamma}(\mu-y_{j})+\sum_{l \in G(j)\setminus \{i\}}\sqrt{\hat{\tau}_{l}}(\tilde{y}_{j}^{(l)}-\hat{b}_{l}-y_{j})]$ and $X_{-i} = (\sqrt{\gamma}+\sum_{l \in G(j)\setminus \{i\}}\sqrt{\hat{\tau}_{l}})$. As before, we substituted $n_{ik} = m_{ik} \cdot \sigma_i$. Since $n_{ik} \sim \mathcal{F}(0,\sigma_{i}^2)$, we get $m_{ik}\sim \mathcal{F}(0,1)$. We see that the absolute value of the above expression is independent of $b_{i}$ and increasing in $\sigma_{i}$. Hence $R(r_j^*,y_j)$ is independent of $b_{i}$ and decreasing in $\sigma_{i}$.

\paragraph{Step 4: $t_{i}^j - \sum_{k \in CG_i^j \setminus \{i\}} w_{ik} t_k^j$ is independent of $b_{i}$ and decreasing in $\sigma_{i}$ for $\sum_{k \in N\setminus\{i\}}w_{ik}\leqslant 1$.}

First, we show that $t_{i}^j - t_{k}^j$ is independent of $b_{i}$ and decreasing in $\sigma_{i}$. This is because $W_j^*$ cancels and this difference reduces to $W_j^{(-k)*}-W_j^{(-i)*}$. The second term is independent of $b_{i}$ and $\sigma_{i}$. The first term is independent of $b_{i}$ and decreasing in $\sigma_{i}$ by the same argument as step 3, with the set of graders reduced to $N \setminus \{k\}$.

Observe that, in the utility of grader $i$, the difference in these two performance score terms appear as follows. \[t_{i} - w_{ik} \cdot \sum_{k \in CG_i \setminus \{i\}} t_{k} = \sum_{j \in NP_i} \left ( t_{i}^j - w_{ik} \cdot \sum_{k \in CG_i^j \setminus \{i\}} t_{k}^j \right ).\]

Consider the terms in the parentheses on the RHS.
\[t_{i}^j - w_{ik} \cdot \sum_{k \in CG_i^j \setminus \{i\}} t_{k}^j = \cdot \sum_{k \in CG_i^j \setminus \{i\}}w_{ik} \cdot (t_{i}^j - t_{k}^j) + (1 - \sum_{k \in CG_i^j \setminus \{i\}}w_{ik}) t_{i}^j.\]
Both terms in the RHS is independent of $b_{i}$ and decreasing in $\sigma_{i}$ as we have already shown and since $\sum_{k \in CG_i^j \setminus \{i\}}w_{ik}\leqslant \sum_{k \in N\setminus\{i\}}w_{ik}\leqslant 1$.
(Note that the first inequality can also be written as $\leqslant \max_{i \in N} \max_{A\subset {\mathcal{F} ^{i}}_{K/2}}\sum_{k \in A}w_{ik}$, since $|CG_i^j| \leqslant K/2 + 1$, which proves the other version of the theorem with a weaker sufficient condition.)

\medskip
Combining all steps, we have shown that the expected utility of grader $i$, where the expectation is taken only w.r.t.\ her true score is independent of $b_{i}$ and decreasing in $\sigma_{i}$. Hence these two properties hold for any choice of actions by the other graders. Hence we have proved that \mechabbrv\ is EPBI and EPRM.

\subsection{Proof of \Cref{thm:grade-unique}}
\label{sec:proof-grade-unique}


Consider the utility expression for agent $i$ with peer-grading performance score weight $\alpha$
\begin{align}
\label{eq:utility-3}
u_{i}(\cdot) &= \max\{r_{i}^{*}(\cdot),y_{i}\}+ \alpha t_{i} - \sum_{j \in G^{-1}(i)} w_{ij} \cdot  \max\{r_{j}^{*}(\cdot),y_{j}\} - \alpha \sum_{k \in CG_i \setminus \{i\}} w_{ik} \cdot t_k \nonumber \\
&= \max\{r_{i}^{*}(\cdot),y_{i}\} - \sum_{j \in G^{-1}(i)} w_{ij} \cdot  \max\{r_{j}^{*}(\cdot),y_{j}\} + \alpha \left ( t_{i} - \sum_{k \in CG_i \setminus \{i\}} w_{ik} \cdot  t_k \right)
\end{align}
As before, the first term on the RHS of the above expression is independent of $\sigma_i$. 
We will show that for all realizations of the random variables $m_{ik}, n_{\ell k}, v_j, \hat{\tau}_\ell$, where $\ell \in G(j) \setminus \{i\}, j \in G^{-1}(i), k \in P_i$, and $\alpha$, the utility is monotone decreasing in $\sigma_i$ if and only if $\lambda_i(\sigma_i) = c_i / \sigma_i$. For brevity of notation, we will use $\lambda_i$ to denote the function where the argument is clear from the context.

We will show that the second term is monotonically decreasing for all realizations of the random variables if and only if $\lambda_i(\sigma_i) = c_i / \sigma_i$. This will complete the proof, since for any other choice of $\lambda_i$, the second term will be non-decreasing (derivative is non-negative) for {\em some} realization of the random variables. The derivative can be zero only when the weights are independent of $\sigma_i$'s. But for that case, the $t_i$'s will also be independent of $\sigma_i$, and hence the third term will also have derivative of zero. When the realizations make the derivative increasing, one can choose $\alpha$ to be small enough such that the decay in the third term is always smaller than the increase in the second term, yielding the overall utility to be increasing in $\sigma_i$.

%

Define the following terms to shorten the forthcoming expressions.
\begin{align*}
 K_1 = \lambda_0 + \sum_{l\in G(j)\setminus\{i\}} \lambda_l, \ K_j = \sum_{l\in G(j)\setminus\{i\}} \lambda_l \left(n_{lj}-\frac{\sum_{k\in P_l}n_{lk}}{x} \right), \ 
 K_{3j} = m_{ij}-\frac{\sum_{k\in P_i}m_{ik}}{x}.
\end{align*}
Consider the second term in \Cref{eq:utility-3}. 
Using \Cref{eq:r-diff}, we reduce the expression for paper $j$ in the sum into the difference term and consider its expectation w.r.t.\ $z_j = \lambda_0 (\mu - y_j)$, which is a $\frac{\lambda_0}{\sqrt{\gamma}} \mathcal{F}(0,1)$ random variable, to get a similar expression like \Cref{expr:step2} as follows. We ignore the positive constant $\frac{\lambda_0}{\sqrt{\gamma}}$ as it does not play a role in determining the sign of the variation.
\begin{equation}
 \label{eq:gen-k}
 I_j = E_{\bar{z}_j \sim \mathcal{F}(0,1)} \max\{r_{j}^{*}(\cdot)-y_{j},0\} = \frac{1}{ K_1 + \lambda_i } \times \int_{0}^{\infty} v_{j} f(v_{j}-K_j-K_{3j} \lambda_i) dv_{j}
\end{equation}
To find the change w.r.t.\ $\sigma_i$, we take its partial derivative and find $\frac{\partial I_j}{\partial \sigma_i}$ to be
\begin{align}
 \frac{
 - K_{3j} \textcolor{blue}{\left( \frac{\partial (\lambda_i \cdot \sigma_i)}{\partial \sigma_i} \right)} ( K_1 + \lambda_i ) \cdot \int_{0}^{\infty} v_{j} f'(v_{j}-K_j-K_{3j} \lambda_i) dv_{j}
 - \textcolor{blue}{\left( \frac{\partial (\lambda_i)}{\partial \sigma_i} \right)} \cdot \int_{0}^{\infty} v_{j} f(v_{j}-K_j-K_{3j} \lambda_i) dv_{j}
 }
 {\left( K_1 + \lambda_i \right)^2}
 \label{eq:derivative}
\end{align}
Note that $K_1$ is positive, while $K_j$ and $K_{3j}$ can take any sign. To ensure that the expression above is positive for {\em all} values of the realized random variables, it is necessary and sufficient that 
\begin{equation}
 \label{eq:nec-suff}
 \frac{\partial (\lambda_i \cdot \sigma_i)}{\partial \sigma_i} = 0, \quad \text{ and } \quad \frac{\partial (\lambda_i)}{\partial \sigma_i} < 0.
\end{equation}
This is because the second integral in the numerator is always positive. Therefore, the above condition is equivalent to $\lambda_i = c_i / \sigma_i$, where $c_i > 0$ is a constant. This concludes the proof.

\section{Calculation of \(r_j^\text{ERM}(\mathbf{\tilde{y}}^{G(j)}_{j}; \hat{\boldsymbol{\theta}}_{G(j)})\) under $\mathbf{PG}_1$~\citep{piech2013tuned} model:}
\label{sec:erm-calculation}

Below, we find a score-assignment function $r_j^\text{ERM}(\mathbf{\tilde{y}}^{G(j)}_{j}, \hat{\boldsymbol{\theta}}_{G(j)})$ that would maximize a quadratic reward function, i.e., that would minimize the expected squared distance between the assigned score and true score on exam $j$.
We will calculate the expression w.r.t.\ the true error parameters $\boldsymbol{\theta}$. Then, given $\boldsymbol{\theta}$ is not observed, we will approximate $\boldsymbol{\theta}$ with the estimated value of the same parameters, i.e., $\hat{\boldsymbol{\theta}}$ to find a new expression.

To calculate $r_j^\text{ERM}(\mathbf{\tilde{y}}^{G(j)}_{j}; {\boldsymbol{\theta}}_{G(j)})$, we first need to calculate the conditional distribution of the true score $y_j$,  $\psi(y_{j}|\tilde{y}_{j}^{G(j)};{b}_{G(j)},{\tau}_{G(j)},\mu,\gamma)$ under the $\mathbf{PG}_1$~\citep{piech2013tuned} model, where $\psi(\cdot)$ is the density of the normal distribution with the mean and variance given by that model. For the convenience of the reader, we restate the $\mathbf{PG}_1$ model below.

\paragraph{Model $\mathbf{PG}_1$ (grader bias and reliability)}
This model puts prior distributions over the latent variables and assumes for example that while an individual grader's bias may be nonzero, the average bias of many graders is zero. Specifically,
\begin{align*}
 \text{(Reliability) } \tau_v &\sim \mathcal{G}(\alpha_0, \beta_0 ) \ \text{ for every grader } v, \\
 \text{(Bias) } b_v &\sim \mathcal{N}(0, 1/\eta_0) \ \text{ for every grader } v, \\
 \text{(True score) } s_u &\sim \mathcal{N}(\mu_0 , 1/\gamma_0) \ \text{ for every user } u, \\
 \text{(Observed score) } z_u &\sim \mathcal{N}(s_u + b_v , 1/\tau_v) \ \text{ for every observed peer grade } s_u,
\end{align*}
where $\mathcal{G}$ and $\mathcal{N}$ refers to the gamma and normal distributions respectively with appropriate hyperparameters. The hyperparameters $\alpha_0, \beta_0,\eta_0, \mu_0 , \gamma_0$ are the hyperparameters for the priors over reliabilities, biases, and true scores, respectively.

Hence, the conditional distribution of the true score $y_j$,  $\psi(y_{j}|\tilde{y}_{j}^{G(j)};{b}_{G(j)},{\tau}_{G(j)},\mu,\gamma)$ is calculated as follows.
\begin{center}
\begin{align*}
    \psi(y_{j}|\tilde{y}_{j}^{G(j)};{b}_{G(j)},{\tau}_{G(j)},\mu,\gamma) &= \frac{\psi(y_{j};\mu,\gamma)\psi(\tilde{y}_{j}^{G(j)}|y_{j};{b}_{G(j)},{\tau}_{G(j)})}  {\int_{y_{j}} \psi(y_{j};\mu,\gamma)\psi(\tilde{y}_{j}^{G(j)}|y_{j};{b}_{G(j)},{\tau}_{G(j)}) dy_j}\\
    &\propto \psi(y_{j};\mu,\gamma)\psi(\tilde{y}_{j}^{G(j)}|y_{j};{b}_{G(j)},{\tau}_{G(j)})\\
    &\propto \psi(y_{j};\mu,\gamma)\prod_{i \in G(j)} \psi(\tilde{y}_{j}^{(i)}|y_{j};{b}_{i},{\tau}_{i})\\
   & \propto \exp \bigg( -\frac{1}{2}\gamma(y_{j}-\mu)^2 + \sum_{i \in G(j)}\Big( -\frac{1}{2}{\tau}_{i}(\tilde{y}_{j}^{(i)} - (y_{j}+{b}_{i}) \Big)^2  \bigg)\\
  &  \propto \exp \bigg( -\frac{1}{2}\Big[\gamma(y_{j}-\mu)^2 + \sum_{i \in G(j)} {\tau}_{i}\big(\tilde{y}_{j}^{(i)} - (y_{j}+{b}_{i}) \big)^2 \Big]  \bigg)
\end{align*}
\end{center}
The expression inside the exponent is quadratic. We consider the exponent as follows.

\begin{center}
\begin{align*}
\gamma(y_{j}-\mu)^2 &+ \sum_{i \in G(j)} {\tau}_{i}\big(\tilde{y}_{j}^{(i)} - (y_{j}+{b}_{i}) \big)^2\\
&= const. + \gamma(y_{j}^{2}-2y_{j}\mu) + \sum_{i \in G(j)} {\tau}_{i}\Big( (y_{j} + {b}_{i} )^{2} - 2\tilde{y}_{j}^{(i)}(y_{j}+{b}_{i})\Big) \\
&= const. + \Big( \gamma + \sum_{i \in G(j)}{\tau}_{i} \Big)y_{j}^{2} - 2\Big( \gamma\mu + \sum_{i \in G(j)}{\tau}_{i}(\tilde{y}_{j}^{(i)}-{b}_{i}) \Big) y_{j},\\
&=const. + R\Bigg( y_{j} - \frac{1}{R} \bigg( \gamma\mu + \sum_{i \in G(j)}{\tau}_{i}(\tilde{y}_{j}^{(i)}-{b}_{i}) \bigg) \Bigg)^{2}\\
&(where\: R= \gamma + \sum_{i \in G(j)} {\tau}_{i})
\end{align*}
\end{center}
Therefore the resultant distribution is Gaussian:
\begin{equation*}
    \psi(y_{j}|\tilde{y}_{j}^{G(j)};{b}_{G(j)},{\tau}_{G(j)},\mu,\gamma) \sim \mathcal{N}\Bigg( \frac{\gamma\mu + \sum_{i \in G(j)}{\tau}_{i}(\tilde{y}_{j}^{(i)}-{b}_{i})}{\gamma + \sum_{i \in G(j)} {\tau}_{i}}, \frac{1}{\gamma + \sum_{i \in G(j)} {\tau}_{i}}\Bigg)
\end{equation*}


\begin{equation}
\label{eqn:expected-score}
\mathbb{E}_{y_{j} | \tilde{y}_{j}^{G(j)};{b}_{G(j)},{\tau}_{G(j)}}\left [ y_{j} \right] = \frac{\gamma\mu + \sum_{i \in G(j)}{\tau}_{i}(\tilde{y}_{j}^{(i)}-{b}_{i})}{\gamma + \sum_{i \in G(j)} {\tau}_{i}}
\end{equation}
Now we are in a position to calculate $r_j^\text{ERM}(\mathbf{\tilde{y}}^{G(j)}_{j}; {\boldsymbol{\theta}}_{G(j)})$.
The reward function is $R(x_{j},y_{j}) = -(x_{j}-y_{j})^{2}$ where $x_{j}$ is the estimated score and $y_{j}$ is the true score for paper $j$. The score-assignment rule {\em expected risk minimizer} (ERM) is given below.
\begin{center}
\begin{align*}
r_j^\text{ERM}(\mathbf{\tilde{y}}^{G(j)}_{j}; {\boldsymbol{\theta}}_{G(j)}) &= \argmax_{x_{j} \in S} \int_{y_{j}} \psi(y_{j}|\tilde{y}^{G(j)}_{j};{b}_{G(j)},{\tau}_{G(j)})R(x_{j},y_{j}) dy_j\\
\text{where } {b}_{i},{\tau}_{i} & \text{ are the estimated bias and reliabilities } \forall i \in G(j)\\
&= \argmax_{x_{j} \in S} \Big[ -\int_{y_{j}} \psi(y_{j}|\tilde{y}^{G(j)}_{j};{b}_{G(j)},{\tau}_{G(j)})(x_{j}-y_{j})^{2}  dy_j \Big]\\
&= \argmin_{x_{j} \in S}\int_{y_{j}} \psi(y_{j}|\tilde{y}^{G(j)}_{j};{b}_{G(j)},{\tau}_{G(j)})(x_{j}-y_{j})^{2} dy_j\\
\end{align*}
\end{center}
Let $g_j(x_{j}) = \int_{y_{j}} \psi(y_{j}|\tilde{y}^{G(j)}_{j};{b}_{G(j)},{\tau}_{G(j)})(x_{j}-y_{j})^{2} dy_j$. Hence we need to find $x_j$ that minimizes $g_j(x_j)$. The first and second order conditions are given as follows.
\begin{center}
\begin{align*}
\frac{\partial g_j(x_{j})}{\partial x_{j}} &= \frac{\partial}{\partial x_{j}}\Bigg[ \int_{y_{j}} \psi(y_{j}|\tilde{y}^{G(j)}_{j};{b}_{G(j)},{\tau}_{G(j)})(x_{j}-y_{j})^{2}  dy_j \Bigg]\\
&= \int_{y_{j}} \frac{\partial}{\partial x_{j}} \Bigg[  \psi(y_{j}|\tilde{y}^{G(j)}_{j};{b}_{G(j)},{\tau}_{G(j)})(x_{j}-y_{j})^{2} dy_j \Bigg]\\
&= 2\int_{y_{j}} \psi(y_{j}|\tilde{y}^{G(j)}_{j};{b}_{G(j)},{\tau}_{G(j)})(x_{j}-y_{j}) dy_j\\
&=2x_{j}\int_{y_{j}} \psi(y_{j}|\tilde{y}^{G(j)}_{j};{b}_{G(j)},{\tau}_{G(j)}) dy_j - 2\int_{y_{j}}y_{j} \psi(y_{j}|\tilde{y}^{G(j)}_{j};{b}_{G(j)},{\tau}_{G(j)}) dy_j\\
&=2x_{j} - 2\mathbb{E}_{y_{j} | \tilde{y}_{j}^{G(j)};{b}_{G(j)},{\tau}_{G(j)}}y_{j}\\
\frac{\partial g_j(x_{j})}{\partial x_{j}} &= 0  \Leftrightarrow x_{j} = \mathbb{E}_{y_{j} | \tilde{y}_{j}^{G(j)};{b}_{G(j)},{\tau}_{G(j)}}y_{j}\\
\frac{\partial^{2} g_j(x_{j})}{\partial x_{j}^{2}} &= 2 > 0 \\
\end{align*}
\end{center}
The first and second order conditions show that $x_{j} = \mathbb{E}_{y_{j} | \tilde{y}_{j}^{G(j)};{b}_{G(j)},{\tau}_{G(j)}}y_{j}$ is a global minima. Hence
\begin{equation}
\label{eqn:erm}
 r_j^\text{ERM}(\mathbf{\tilde{y}}^{G(j)}_{j}; {\boldsymbol{\theta}}_{G(j)}) = \mathbb{E}_{y_{j} | \tilde{y}_{j}^{G(j)};{b}_{G(j)},{\tau}_{G(j)}}y_{j} = \frac{\gamma\mu + \sum_{i \in G(j)}{\tau}_{i}(\tilde{y}_{j}^{(i)}-{b}_{i})}{\gamma + \sum_{i \in G(j)} {\tau}_{i}}.
\end{equation}
The last equality follows from \Cref{eqn:expected-score}.

Replacing $\boldsymbol{\theta}$ with the estimated parameters, i.e., $\hat{\boldsymbol{\theta}}$, we get,
\begin{equation}
\label{eqn:erm_est}
 r_j^\text{ERM}(\mathbf{\tilde{y}}^{G(j)}_{j}; \hat{\boldsymbol{\theta}}_{G(j)}) = \mathbb{E}_{y_{j} | \tilde{y}_{j}^{G(j)}; \hat{b}_{G(j)}, \hat{\tau}_{G(j)}}y_{j} = \frac{\gamma\mu + \sum_{i \in G(j)}{\hat\tau}_{i}(\tilde{y}_{j}^{(i)}-\hat{b}_{i})}{\gamma + \sum_{i \in G(j)} {\hat\tau}_{i}}.
\end{equation}

\section{Comparison of $W_j^{\text{ERM}}$ and 
$W_j^{\text{ISWDM}}$}
\label{sec:comparison2}

\begin{figure*}[h!]
        \begin{subfigure}[b]{0.49\linewidth}
                
\includegraphics[width=1.0\linewidth]{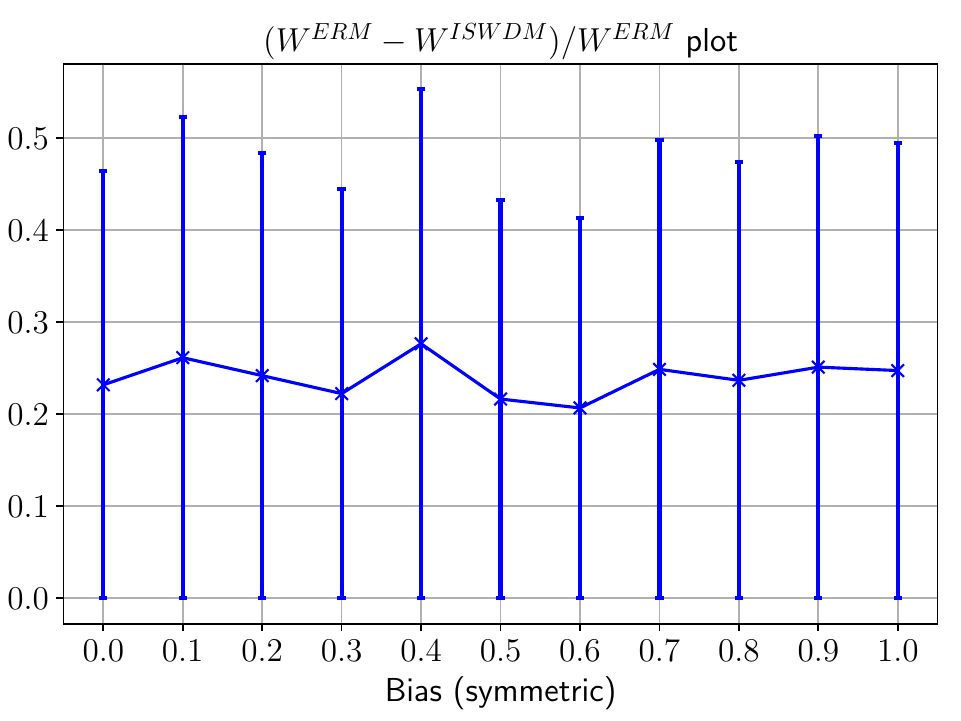}
        \end{subfigure} %
        \begin{subfigure}[b]{0.49\linewidth}
                
\includegraphics[width=1.0\linewidth]{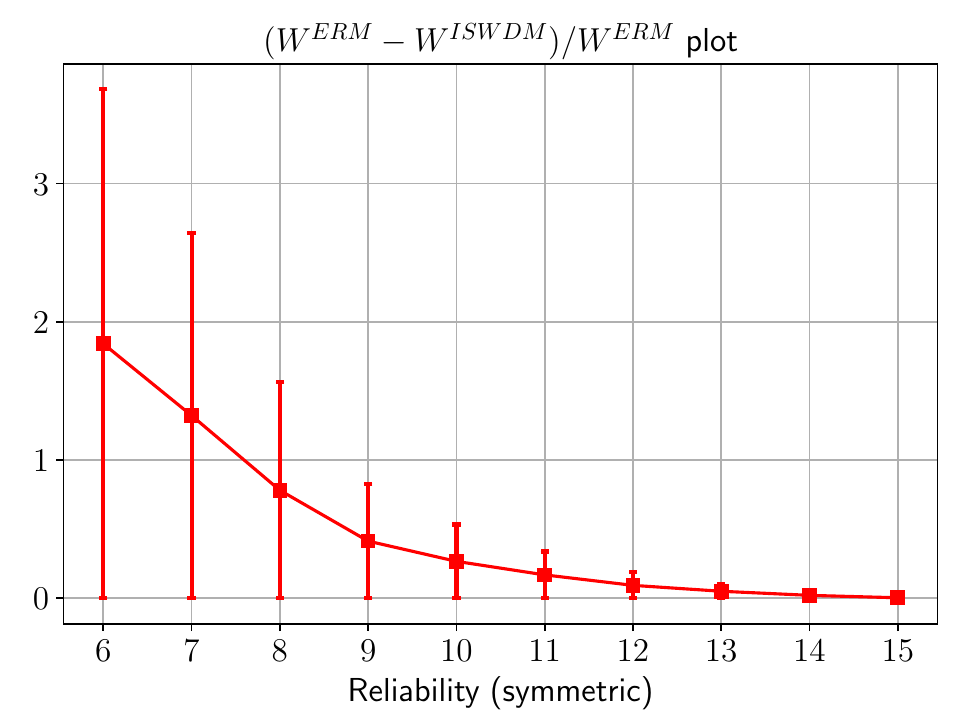}
        \end{subfigure}%
        \caption{Sub-optimality with increasing bias and 
reliability}
        \label{bias_rel_error}
\end{figure*}

\Cref{bias_rel_error} shows the sub-optimality, i.e., 
$W_j^{\text{ERM}}-W_j^{\text{ISWDM}}$ as a fraction of $W_j^{\text{ERM}}$, with respect to an 
increasing symmetric bias and reliability, where `symmetric' 
means that every peer-grader has the same bias and 
reliability. We ran the simulation for the peer-grading model proposed by \citet{piech2013tuned} with parameters $\mu=1, \gamma=16$, and the reward function $R(r,y) = -(r-y)^2$. The simulation is repeated $100$ times for each bias and reliability to obtain the statistical measures. 
It shows that the sub-optimality is small but insensitive 
to bias (roughly $20\%$) and monotonically decreasing in reliability.

\section{Instructions provided to the human subjects (\Cref{sec:human})}
\label{sec:instructions}

The instructions for both the mechanisms were as follows.

\subsection{Median Mechanism Instructions}

First, please register yourself on: [registration link] and solve the problem\footnote{The problem tests the participant's understanding of median and similar simple techniques.} therein. The example there would help you understand how your decisions map into your final payments, through the median-mechanism and payment system used in this study.
This is a study on peer-grading. You should read the following instructions carefully, as they would help you perform successfully in the study. In this study, each of you will be asked to grade the assignments of five anonymous students in this room. Similarly, your own assignment would be graded by five anonymous students from this room.
Your peer-graded marks and the relative ranks in this peer-grading exercise only determine your payment from this session. It will not be used to determine your actual score for your final grade in the course. The assignment score used towards your university grades will be provided to you by the instructor (i.e., tutors or myself) later.

\paragraph{Would I know whose exam papers I might be grading / correcting?}
You would not have this information. We will take maximal precautions to make sure that the grader or the assignment-owner's identities are anonymous to each other during and after this session. Further you would also not know which other four participants are grading the same papers as you. Thus, this procedure is double-blind.
We will provide you a solution manual to help you in the grading process. Follow the explanation of the questions and correct answers presented before the study. Please be respectful and encouraging in the grading process. Scores should reflect the learner's understanding of the assignment and points should not be deducted for difficulties with language or differences in opinion or for using a different but correct methodology.

How are the final grades on my own assignment decided?
All five peer-graders independently assign you grades on all of the questions (there are 5 in total, all worth 2 points). Then for each question-part your final grade is the median of those five grades. For example, if on the second round of peer-grading, the five graders assign you 0, 1, 1.5, 2, and 2 respectively, then your final assigned grade on that question would be 1.5.
We would calculate your grades on all the questions separately by the above median-method, and then aggregate those median grades from all the questions. For example, if there are five questions and the median grades on the questions are 0, 1, 1.5, 2 and 2 respectively, then the total grade on the assignment is 6.5.

\paragraph{How does one calculate the median of five numbers?}
Sort the numbers in increasing order and the third highest number would be the median.

\paragraph{Can I dispute my peer-assigned grades?}
Yes, for certain questions you can, and for others you cannot. In case you think your true grade is different than the grade that has been assigned to you on these questions, you can privately indicate that on a form, that would be sent at the end of the peer-grading and that will immediately notify us. We would then reassign you the grade the Teaching staff had assigned to your assignment previously. This whole process would be completed in a click of a button and you would be shown your updated grade in a matter of seconds. 
Please note that once a dispute is lodged, your grade would become the Teaching Staff assigned grade irrespective of whether that results in an increase or decrease over your original grade.

\paragraph{How are my payments decided?}
Every participant would get a show up fee of M 50 for participating in and completing this session. You would also get an additional amount depending on your ranking in the pool of `n' participants today. The ranking would be done in decreasing order of the final grades assigned to you all on the whole assignment. A ranking of x means that there are (x-1) other people who have a strictly higher grade than you. The additional amount would be equal to M 650 for the top 25\% (first quartile) ranked students, M 450 for the next 25\% (second quartile) ranked students, M 250 for the third quartile ranked students, and M 50 for the bottom quartile students.
If the number of students that scored the same overlaps to two or more different quartiles, then all of them get the average payment of those quartiles. E.g., suppose 7 students got the same marks, and 3 students are in first quartile while 4 are in second quartile -- then all 7 get M 600 (average of M 700 and M 500).
Hence, in this study, the higher you are in the ranking based on your peers' judgment (and a potential review), higher is your total payment.

\paragraph{How do the grades you submit affect your own payment?}
The grades you submit obviously do not affect your own grade, because you are never grading your own paper, but they can still affect your own payment. Your grading would potentially affect the grades of others, and that can change the relative rank between you and the person(s) you are grading. For example, when you assign someone a higher grade, that might change the median grade they are assigned, and thus move them to a higher rank than you. Similarly, when you give them a lower grade, it might move them to a relatively lower rank than you. Obviously, both of these scenarios would affect the final payments of both you and the other person, as everyone is paid according to the final rankings.

Time-line for the study in chronological order:
\begin{itemize}
 \item Stage 0: The whole assignment to be graded is broken up into 3 small parts, that would be peer-graded in three stages. The total grade from the whole assignment determines your final ranking and payment. At this stage, you are expected to complete the questionnaire successfully.
 \item Stage 1: Every one of you peer-grades the first part of the assignment of 5 of your peers. Therefore, for any question you are grading in this stage, you know that 4 other anonymous participants are also grading that question. Also, the first part of your own assignment is also being peer-graded by 5 other participants. One part of these questions will have options for regrading, while the other part will not (it will be mentioned in the response sheet, but all regrading requests will be collected at the end of stage 3).
 \item Feedback Stage 1: For each paper you graded in Stage 1, we will show you the grades assigned by you and the 4 other anonymous graders. We will also show you how part 1 of your own assignment got graded by the assigned graders.
 \item Stage 2: Similar to Stage 1, now part 2 of the assignment gets peer-graded. But the papers are now sent to a new random set of peer-graders. One part of these questions will have options for regrading, while the other part will not (it will be mentioned in the response sheet, but all regrading requests will be collected at the end of stage 3).
 \item Feedback Stage 2: Feedback of Stage 2 (similar to Stage 1) observed.
 \item Stage 3: Similar to Stage 2 (one part has regrading requests, the other does not), now part 3 of the assignment gets peer-graded.
 \item Feedback Stage 3: Feedback of Stage 3 (similar to Stages 1 and 2) is sent to all students, along with their tentative total score. You may raise regrading requests for the part that is regradable (as mentioned above). Any regrading requests that are lodged will be acted on. Performance on the whole assignment is aggregated, and the final ranking and payments are sent via email. To finish the study, complete the survey that comes in the last email.
Study ends.
\end{itemize}

\paragraph{Is my data confidential?}
Yes, your data is completely confidential. Before observing and analyzing the collected data, we would be removing every personal identifier from the data, so that none of the decisions can be traced back to the individual who made the decision.

The first practice example tests you on your understanding of the mechanism how the peer-grading leads to your final grade, rank, and payment. You must complete this practice example with a score of 80\% or more (i.e., correctly answer at least 4 questions out of 5).  You will get one chance only, so please do this carefully. Failing this, you would be asked to leave this session with a M 20 reward.

{\bf Important: Please do not communicate with any other participants during this session. For the grading, open one file at a time, finish grading, submit the grade in the google form and then move on. Please keep seated even if you are done with grading before time. If you have any questions, please raise your hand and one of us will come by to answer your query. Please use your university domain email id throughout this session. Please come remembering your google id/password, since that may be needed for some form filling.}

\subsection{\mechabbrv\ Instructions}

Before you begin, please register yourself on: [registration link]. Submit the form
only once. 

This is a study on peer-grading. In this study, each of you will be asked to grade {\bf five} anonymous assignments. Similarly, your
own assignment would be graded by a certain number of anonymous students from this
room. Your peer-graded marks and your performance in the peer-grading
exercise will only determine your payment from this session. It will
not be used to determine your actual score for your final grade in
the course. The assignment score used towards your university grades will
be provided to you by the instructor (i.e., tutors or myself) later.

\paragraph{Would I know whose exam papers I might be grading / correcting? }

You would not have this information. We will take maximal precautions
to make sure that the grader or the assignment-owner's
identities are anonymous to each other during and after this session.
Further you would also not know which other four participants are
grading the same papers as you. Thus, this procedure is double-blind.
We will provide you a solution manual to help you in the grading process.
Follow the explanation of the questions and correct answers presented
before the study. Please be respectful and encouraging in the grading
process. Scores should reflect the learner's understanding
of the assignment and points should not be deducted for difficulties
with language or differences in opinion or for using a different but
correct methodology.

\paragraph{How are the final grades on my own assignment decided? }

Your peer-graders independently assign you grades on all of the
questions. Then for each question your final grade is decided by running
it through a {\bf new mechanism called \mechabbrv\ (\mech)}. This is a mechanism which is designed
to remove the individual biases in grading, and selectively weight
and reward graders by how precise they are (details to follow). We would
calculate your grades on all the questions separately by the above
method, and then aggregate those grades from all the questions. In each round, you will have some {\bf regradable} and {\bf non-regradable} questions. For the regradable part, you will earn the peer-given score computed through \mechabbrv\ {\bf and} an additional \mechabbrv\ reward for grading. For the non-regradable part, you will only receive the peer-given score computed through \mechabbrv, {\bf but no additional reward for grading}.

\paragraph{Can I dispute my peer-assigned grades?} 

Yes, for certain questions you can, and for others you cannot. In
case you think your true grade is different than the grade that has
been assigned to you on these questions, you can privately indicate
that on a form, that would be sent at the end of the peer-grading
and that will immediately notify us. We would then reassign you the
grade the Teaching staff had assigned to your assignment previously.
This whole process would be completed in a click of a button and you
would be shown your updated grade in a matter of seconds. Please note
that once a dispute is lodged, your grade would become the Teaching
Staff assigned grade irrespective of whether that results in an increase
or decrease over your original grade.

\paragraph{What is the \mechabbrv\ mechanism?} 

Let us describe \mechabbrv\ in short in the following two steps: \\
\textbf{Step 1 Probes:} Out of the five questions you (a grader) grade,
two are randomly assigned to be \textit{probes} (rest three are \textit{non-probes}).
On the probe papers, we would directly assign the teaching staff assigned
grades and also use the teaching staff assigned grades to get an estimate
of your individual \textit{average deviation (or bias)} and \textit{variance}
in the assignments you graded. We will do this for all the graders.
For a grader who on average, assigns a grade higher than the true-grade,
the estimated deviation would be negative, and otherwise would be
positive.\\
\textbf{Step 2 Non-Probes:} The non-probes would be graded using the
information from, (i) the assigned grades of all the graders, and
(ii) the estimated average deviation (or bias) and variance of grading
by peer-graders in Step 1. The assigned scores would be ``de-biased''
using the information in 2.\\
Here is a numerical example that goes through these two steps. Suppose
on the five questions you graded, the first two questions are randomly
assigned as probes (this is for illustration only, the actual probes will be interspersed and not the first two, and you won't know which are the probes).
\begin{center}
\begin{tabular}{|c|c|c|c|c|c|}
\hline 
\textcolor{gray}{Paper} & \textcolor{gray}{Status} & \textcolor{gray}{Score you} & \textcolor{gray}{True} & \textcolor{black}{Deviation} & \textcolor{black}{Bias=Avg of}\tabularnewline
 &  & \textcolor{gray}{assigned (A)} & \textcolor{gray}{Score (B)} & \textcolor{black}{(A-B)} & \textcolor{black}{Deviation}\tabularnewline
\hline 
\hline 
\textcolor{gray}{1} & \textit{\textcolor{gray}{Probe}} & \textcolor{gray}{3} & \textcolor{gray}{3} & \textcolor{black}{3-3=0} & \multirow{2}{*}{\textcolor{black}{$\frac{0+(-.5)}{2}=-.25$}}\tabularnewline
\cline{1-5} 
\textcolor{gray}{2} & \textit{\textcolor{gray}{Probe}} & \textcolor{gray}{2.5} & \textcolor{gray}{2} & \textcolor{black}{2-2.5=-.5} & \tabularnewline
\hline 
\textcolor{gray}{3} &  & \textcolor{gray}{3.5} &  &  & \tabularnewline
\hline 
\textcolor{gray}{4} &  & \textcolor{gray}{4} &  &  & \tabularnewline
\hline 
\textcolor{gray}{5} &  & \textcolor{gray}{2} &  &  & \tabularnewline
\hline 
\end{tabular}
\par\end{center}

On the probe questions, your evaluation would be compared with the
evaluation done by the course instructors (True score), to calculate
an average deviation in your grading. We would then use this to calculate
the variance of your deviation.
\begin{center}
\begin{tabular}{|c|c|c|c|c|c|c|}
\hline 
\textcolor{gray}{Paper} &  & \textcolor{gray}{Score you} & \textcolor{gray}{True} & \textcolor{gray}{Deviation} & \textcolor{gray}{Bias=Avg} & \textcolor{black}{Variance of}\tabularnewline
 &  & \textcolor{gray}{assigned} & \textcolor{gray}{Score} &  & \textcolor{gray}{Deviation} & \textcolor{black}{Deviation}\tabularnewline
\hline 
\hline 
\textcolor{gray}{1} & \textit{\textcolor{gray}{Probe}} & \textcolor{gray}{3} & \textcolor{gray}{3} & \textcolor{gray}{3-3=0} & \multirow{2}{*}{\textcolor{gray}{$-.25$}} & \textcolor{black}{$\frac{(0+.25)^{2}+(-.5+.25)^{2}}{2}$}\tabularnewline
\cline{1-5} 
\textcolor{gray}{2} & \textit{\textcolor{gray}{Probe}} & \textcolor{gray}{2.5} & \textcolor{gray}{2} & \textcolor{gray}{2-2.5=-.5} &  & \textcolor{black}{$=.0625$}\tabularnewline
\hline 
\textcolor{gray}{3} &  & \textcolor{gray}{3.5} &  &  &  & \tabularnewline
\hline 
\textcolor{gray}{4} &  & \textcolor{gray}{4} &  &  &  & \tabularnewline
\hline 
\textcolor{gray}{5} &  & \textcolor{gray}{2} &  &  &  & \tabularnewline
\hline 
\end{tabular}
\par\end{center}

Suppose the (bias,variance) pairs of the other two graders, who are
also grading question 4, are \textcolor{blue}{(.25, .05)} and \textcolor{red}{(-.5,.2)}
respectively. Suppose the scores they had assigned to the same Q4 was 3 and 2 respectively, while you have given 4 to that question.

Then, the final grade on Q4 (a typical non-probe question)
would be calculated as ($k_1$ and $k_2$ are some appropriately chosen constants)

\textcolor{black}{
\begin{eqnarray*}
\text{assigned\_score} & = & {\color{black}\frac{k_{1}+{\color{gray}\frac{1}{\sqrt{.0625}}(4+(-.25))}+{\color{blue}\frac{1}{\sqrt{.05}}(3+.25)}+{\color{red}\frac{1}{\sqrt{.2}}(2+.5)}}{k_{2}+{\color{gray}\frac{1}{\sqrt{.0625}}}+{\color{blue}\frac{1}{\sqrt{.05}}}+{\color{red}\frac{1}{\sqrt{.2}}}}}
\end{eqnarray*}
}When we assign the final grade on any non-probe question, we will
\textquotedblleft {\bf de-bias}\textquotedblright{} the reports from all
the graders by subtracting out the bias, and also selectively over-weight
the information from the low-variance graders. We consider the inverse
of the square-root of your variance as your {\bf precision of grading}, and
use this precision to weight your assigned score on this paper. The
accuracy of the mechanism\_assigned\_score is given by $-\text{(assigned\_score-true\_score)}^{2}$.

If you were not one of the graders, and the mechanism only assigned
scores using the reports of the other graders,

\textcolor{black}{
\begin{eqnarray*}
\text{assigned\_score\_without\_you} & = & \frac{k_{1}+{\color{blue}\frac{1}{\sqrt{.05}}(3+.25)}+{\color{red}\frac{1}{\sqrt{.2}}(2+.5)}}{k_{2}+{\color{blue}\frac{1}{\sqrt{.05}}}+{\color{red}\frac{1}{\sqrt{.2}}}}
\end{eqnarray*}
}

The new accuracy is $-\text{(assigned\_score\_without\_you-true\_score)}^{2}$.
Now, your \mechabbrv\ performance score from peer-grading question 4 would be calculated
as the difference between the accuracy with you, and the accuracy
without you. This is intuitively equivalent to you getting paid for
your relative contribution in your group towards making the final
assigned grade accurate. {\bf The more accurate the assigned score is,
when you are included in the group of graders, the higher would be
your performance score!}

The \mechabbrv\ performance score on each question you have graded that is worth
$x$ points, is assigned on the scale of $[0,\frac{x}{2}]$. So, in
round 1, where each regradable question is worth one point, and you
grade a total of 3 non-probe questions, the maximum \mechabbrv\ performance score
you could get is $3\times 0.5 =1.5$ and the minimum is $0$.

This \mechabbrv\ grade and performance scores have the following properties: 

\medskip
\textbf{Bias Invariance:} Suppose you had reported grades of 3+x,
2.5+x, 3.5+x, 4+x, and 2+x, on all the questions instead, and thus
had an individual deviations x points higher than before. This would
have no effect on the \mechabbrv\ performance scores, as it would be de-biased as
described above. This is a mathematical property of the mechanism
described. 

With the new reported grades, your average deviation is changed to
$-.25-x$ from $-.25.$ The $+x$ and $-x$ cancel out in the expression
of the assigned score, leaving it unchanged.

\textcolor{black}{
\begin{eqnarray*}
\text{assigned\_score} & = & {\color{black}\frac{k_{1}+{\color{gray}\frac{1}{\sqrt{.0625}}(4+\cancel{x}+(-.25-\cancel{x}))}+{\color{blue}\frac{1}{\sqrt{.05}}(3+.25)}+{\color{red}\frac{1}{\sqrt{.2}}(2+.5)}}{k_{2}+{\color{gray}\frac{1}{\sqrt{.0625}}}+{\color{blue}\frac{1}{\sqrt{.05}}}+{\color{red}\frac{1}{\sqrt{.2}}}}}
\end{eqnarray*}
}

{\bf Clearly the assigned\_score\_without\_you also cannot change if your
bias changes, so your expected \mechabbrv\ performance score cannot change here!}

\medskip
\textbf{Precision Monotonicity:} For every set of (bias, variance)
your co-graders might have, your expected \mechabbrv\ performance score from the
peer-grading task is monotonically increasing in your grading-precision
(\textit{precision is the inverse square-root of your variance}). This is a mathematical
property that can be easily showed by using calculus and statistics.
Thus the more precisely you evaluate a paper in the peer-grading task,
(or alternatively the lower your grading variance) the higher your
peer-grading score. 

Here is a graph that shows how the \mechabbrv\ performance score changes with the
Precision for a grader, who is grading alongside with two graders,
one of highest precision and one of lowest precision.


\paragraph{How do you calculate $\boldsymbol{-\text{(assigned\_score-true\_score)}^{2}}$?} 

If there is no regrading request, then we would assume that true\_score=assigned\_score,
and this value is zero. If there are regrading requests, then we would
evaluate the paper ourselves and assign the course-instructor assigned
score as true\_score to calculate the value.

\paragraph{What is my consolidated score?} 

Your consolidated score is the sum of (i) the score on your own assignment (consolidated score from the regradable and non-regradable parts),
and (ii) your \mechabbrv\ performance score (peer-grading score). For example,
if the peer-assigned score (computed via \mechabbrv) on your own assignment is x, and your peer-grading
score is y, your consolidated score is x+y.

\paragraph{How are my payments decided?} 

Every participant would get a show-up fee of M 50 for participating
in and completing this session. You would also get an additional amount
depending on your ranking in the pool of \textquoteleft n\textquoteright{}
participants today, based on the consolidated score. {\bf The ranking would
be done in decreasing order of the final grades (i.e., the consolidated score) assigned to you all
on the whole assignment}. A ranking of x means that there are (x-1)
other people who have a strictly higher consolidated score than you.
The additional amount would be equal to M 650 for the top 25\% (first
quartile) ranked students, M 450 for the next 25\% (second quartile)
ranked students, M 250 for the third quartile ranked students, and
M 50 for the bottom quartile students. If the number of students
that scored the same overlaps to two or more different quartiles,
then all of them get the average payment of those quartiles. For example,
suppose 13 students out of a population of 40 got 10/10, then all 13 get M $(700 \times 10 + 500 \times 3)/13 = 654$ -- the next rank starts from 14. Hence, in this study, the higher is
your consolidated score, higher is your total payment.

\paragraph{How do the grades you submit affect your own payment?} 

The grades you submit obviously do not affect your own grade, because
you are never grading your own paper, but they can still affect your
own payment, in two ways. \\
1) {\bf By affecting the grade of others}: Your grading could potentially
affect the grades of others, \textit{only if the question is chosen
as non-probe question}, and consequently that can change the relative
rank between you and the person(s) you are grading. For example, when
you assign someone a higher/ lower grade on a question that is chosen
as a non-probe question, that might change the \mechabbrv\-assigned quiz
score (and thus the consolidated score) they are assigned, and thus
affect the relative rankings. But, note that Bias Invariance result
described above already tells you that a \textbf{different bias would
not change the expected quiz scores} of any peers. \\
2) {\bf By affecting your peer-grading score}: Assigning a higher/ lower
score on any question, could change your payments in two ways. If
this happened on a question that was chosen as probe, we would be
calculating your precision and bias to a different number, and \textbf{a
lower (respectively higher) precision would result in a lower (respectively
higher) marginal impact of your peer-grading reports, and hence, a
lower (respectively higher) peer-grading score (and hence lower consolidated
score)} for you. If this was a non-probe question instead, then you
might be able to change the peer-graded score on that paper, depending
on how much weight we assign to your evaluation.

\paragraph{Is my data confidential?} 

Yes, your data is completely confidential. Before observing and analyzing
the collected data, we would be removing every personal identifier
from the data, so that none of the decisions can be traced back to
the individual who made the decision.

You would be given a questionnaire of three questions that tests you
on your knowledge of calculation of median. Failure in answering at
least two correctly out of those three questions would disqualify
you from participation in this study. In this case you would be asked
to leave this session with a M 20 reward. Important: Please do not
communicate with any other participants during this session. For the
grading, open one file at a time, finish grading, submit the grade
in the google form and then move on. Please keep seated even if you
are done with grading before time. If you have any questions, please
raise your hand and one of us will come by to answer your query. Please
use your university domain email id throughout this session. Please come
remembering your google id/password, since that may be needed for
some form filling.

\end{document}